\documentclass[12pt]{iopart}
\usepackage{epsfig,xspace}

\def\nue{\ensuremath{\nu_{e}}\xspace}

\def\numu{\ensuremath{\nu_{\mu}\xspace}}

\def\nutau{\ensuremath{\nu_{\tau}\xspace}}

\newcommand{\nuenumu}{\ensuremath{\nue \rightarrow \numu\xspace}}
\newcommand{\numunutau}{\ensuremath{\numu \rightarrow \nutau\xspace}}
\newcommand{\nuenutau}{\ensuremath{\nue \rightarrow \nutau}}

\newcommand{\dmot}{\ensuremath{\Delta m^2_{21}\xspace}}
\newcommand{\dmtt}{\ensuremath{\Delta m^2_{32} \xspace}}
\newcommand{\ldm}{\Delta m_{31}^2}
\newcommand{\sdm}{\Delta m_{21}^2}

\def\Li{^6{\mathrm{Li}}}
\def\Li8{\ensuremath{^8{\mathrm{Li}}}}
\def\B8{\ensuremath{^8{\mathrm{B}}}}
\def\anue{\overline{{\mathrm\nu}}_{\mathrm e}}
\def\anumu{\overline{{\mathrm\nu}}_{\mathrm \mu}}
\newcommand{\thetaot}{\ensuremath{\theta_{13}}\xspace}
\newcommand{\thetatt}{\ensuremath{\theta_{23}}\xspace}
\newcommand{\numunue}{\ensuremath{\nu_\mu \rightarrow \nu_e}\xspace}

\newcommand{\be}{\begin{equation}}
\newcommand{\ee}{\end{equation}}
\newcommand{\bea}{\begin{eqnarray}}
\newcommand{\eea}{\end{eqnarray}}

\begin{document}

\title[Learning from tau appearance]{Learning from tau appearance}

\author{P Migliozzi}
\address{ I.N.F.N., Sezione di Napoli, Naples, Italy }

\author{F Terranova}
\address{ I.N.F.N., Laboratori Nazionali di Frascati,
Frascati (Rome), Italy }

\begin{abstract}
The study of \numunutau\ oscillation and the explicit observation of
the \nutau\ through the identification of the final-state tau lepton
(``direct appearance search'') represent the most straightforward
test of the oscillation phenomenon. It is, nonetheless, the most
challenging from the experimental point of view. In this paper we
discuss the current empirical evidence for direct appearance of tau
neutrinos at the atmospheric scale and the perspectives for the next
few years, up to the completion of the CNGS physics programme.  We
investigate the relevance of this specific oscillation channel to
gain insight into neutrino physics within the standard three-family
framework. Finally, we discuss the opportunities offered by
precision studies of \numunutau\ transitions in the occurrence of more
exotic scenarios emerging from additional sterile neutrinos or
non-standard interactions.
\end{abstract}

\maketitle

\section{Introduction}
\label{sec:intro}

The search for neutrino oscillations into flavors different from the
initial one (``appearance'') has a decades-long history.  Since 1998,
however, the study of the $\numunutau$ transition has played a unique
role in the field of neutrino physics for a very special
reason. 

Neutrino oscillations~\cite{giunti_book} are a powerful tool to
determine the squared mass differences of the rest masses of neutrinos
because the oscillation phase is proportional to the ratio $(m_i^2 -
m_j^2)L/E \equiv \Delta m_{ij}^2 L/E$.  Here, $i$ and $j$ label the
mass eigenstates ($i,j=1,2,3$), $L$ the source-to-detector distance
(``baseline'') and $E$ the neutrino energy. Oscillations also depend
on the $3 \times 3$ matrix that describes the mismatch of flavor and
mass eigenstates. It is the leptonic counterpart of the
Cabibbo-Kobayashi-Maskawa matrix and is often referred to as the
Pontecorvo-Maki-Nakagawa-Sakata (PMNS) matrix
\cite{pontecorvo,neutrino_osc}. This matrix can be parameterized by
three angles, $\theta_{12},\theta_{23}$ and $\theta_{13}$, and one
CP-violating phase $\delta$~\footnote{Additional Majorana phases
cannot be observed by oscillation experiments~\cite{giunti_book}.}.  The experimental
results obtained so far point to two very distinct mass differences,
$\Delta m^2_{sol} = \dmot \equiv m^2_2 -m^2_1 = 7.65^{+0.23}_{-0.20}
\times 10^{-5}$ eV$^2$~\cite{SchwetzTortolaValle} and $|\Delta m^2_{atm}| = |\dmtt| \equiv |m^2_3
-m^2_2| \simeq |m^2_3 -m^2_1| = 
2.32^{+0.12}_{-0.08} \times 10^{-3} \
\mathrm{eV}^2$~\cite{minos_adamson}.
\dmot\ is called the ``solar
mass scale'' because it drives oscillation of solar neutrinos, but, of
course, if the energy of the neutrino and the source-to-detector
distance are properly tuned, it can be measured also employing man-made
neutrinos, e.g., reactor neutrinos located about 100~km from the
detector~\cite{kamland}. Similarly, atmospheric neutrinos mainly
oscillate at a frequency that depends on \dmtt (``atmospheric
scale''). Accelerator neutrino experiments can see (actually, saw in
K2K~\cite{K2K} and MINOS~\cite{minos}) the same effect using neutrinos
of energy $\mathcal{O}(1)$~GeV and baselines of a few hundreds of km.

Appearance has always been considered the most direct proof of
the phenomenon of neutrino oscillation. Unfortunately, all sources
that we have at our disposal to observe oscillations at the solar
scale  (solar and
reactor neutrinos) produce $\nue$ (or $\anue$) with energy well below
the kinematic threshold for muon production. As a consequence, it is
impossible to test in a straightforward manner the occurrence of $\nue
\rightarrow \numu$ or $\nue \rightarrow \nutau$ transitions through
the observation of muons or taus produced by charged-current (CC)
neutrino interactions with matter. At the atmospheric scale
(atmospheric and multi-GeV artificial neutrinos from the decay in
flight of pions), \numunue transitions might be observed in appearance
mode. Still, the peculiar structure of the leptonic mixing matrix
suppresses this transition at least by one order of magnitude by
virtue of the small \thetaot
angle~\cite{chooz,palo_verde,Mezzetto:2010zi}. Therefore, an
appearance measurement that is aimed at observing a large
(i.e. $\mathcal{O}$(1)~) neutrino transition probability must resort
to \numunutau. In the current framework of interpretation of neutrino
oscillation data - three active neutrinos non-trivially mixed by a
$3\times3$ unitary matrix~\cite{PDG} - such probability is quite large
for multi-GeV neutrinos at baselines of the order of $10^{3}$~km. In
this case, the oscillation probability is given by:
\be
P(\numunutau) \simeq \cos^4 \thetaot \sin^2 2 \thetatt \sin^2 \Delta_{32}
\ee
$\Delta_{32} = \dmtt L/4E$ being the oscillation phase (L is the
baseline and E the neutrino energy in $c=\hbar=1$ units) while
\thetaot and \thetatt are the mixing angles of the third family with
the first and second one, respectively. Due to the tri-bimaximal
structure~\cite{tribimaximal} of the mixing matrix ($\thetaot \simeq 0$
and $\thetatt \simeq \pi/4$), this probability is ${\mathcal O}$(1) at
the oscillation peak ($\Delta_{32} = \pi/2$).

Seeking for \numunutau, i.e. observing final state $\nutau$ CC
interactions, is a major experimental challenge. The source must
produce neutrinos well above the kinematic threshold for tau
production (3.5~GeV for scattering in nuclei). Moreover, the far detector has to be capable of
selecting an enriched sample of tau leptons in the bulk of muons and
hadrons produced by $\numu$ CC and NC interactions. It comes as no
surprise that the most direct test of the oscillation phenomenon
through the observation of tau appearance still deserves a conclusive
evidence. In 2010, however, important milestones have been achieved,
especially by the SuperKamiokande and OPERA experiments. The aim of
this paper is a careful examination of the present evidence for tau
appearance as a direct probe of neutrino oscillations.  In addition, we
discuss what can be learned from \numunutau\ studies in the standard
3-family oscillation framework and in more exotic scenarios. We also
anticipate the relevance of precision measurements of \numunutau\ to
be performed by a future generation of short/long baseline
experiments.

%
\section{Past searches for tau appearance}
\label{sec:early}
At the beginning of the 90's, there were theoretical
arguments~\cite{Harari:1988ew, Ellis:1992zr} suggesting that, in
analogy with quark mixing, neutrino mixing angles should be small and
that the heavier neutrino (mostly $\nu_\tau$) may have a mass of 1 eV,
or larger, and therefore could be the main constituent of the dark
matter in the universe.  This hypothesis was based on two key
assumptions:
\begin{itemize} 
\item the interpretation of the solar neutrino deficit in
terms of $\nu_{e} \to \nu_{\mu}$ oscillations amplified by matter
effects, giving $\Delta m^{2} \approx 10^{-5}$ eV$^{2}$;
\item the input from see-saw mass-generation models~\cite{seesaw}, which predicts that
neutrino masses are proportional to the square of the mass of the
charged lepton, or of the  2/3 charge quark of the same family.

\end{itemize} 
From these two assumptions one expects a $\numu \simeq \nu_2$ mass of $\sim 3\times
10^{-3}$ eV and a $\nu_\tau \simeq \nu_3$ mass of $\sim 1$ eV, or
higher. 

The seeming concordance between cosmological and particle physics
hints boosted enormously the search for tau appearance; in particular,
it supported the design and construction of two high sensitivity short
baseline experiments to discover  $\nu_\mu\rightarrow
\nu_\tau$ oscillations in the region of $\Delta m^2\sim
10$~eV$^2$. The two CERN experiments performing this search were
NOMAD~\cite{Altegoer:1997gv} and CHORUS~\cite{Eskut:1997ar}, both exploiting the CERN SPS wide-band neutrino
beam (WANF~\cite{wanf}) but with two
quite different approaches. 

%
\subsection{The NOMAD experiment}
\label{expe_nomad}

The NOMAD experiment was designed to search for
$\nu_\mu\rightarrow\nu_\tau$ in the WANF. 
The detector consisted of drift chambers used as target and tracking
medium. They were optimized to fulfill two opposite requirements: a
heavy target to collect as many interactions as possible and a light
target to allow a precise tracking by reducing multiple scattering. In
total there were 44 chambers with a fiducial mass of 2.7~tons and an
active area of 2.6 $\times$2.6 m$^2$. They were followed by a
transition radiation detector (TRD) for e/$\pi$ separation. Electron
identification was performed with a pre-shower detector and a
lead-glass electromagnetic calorimeter followed by an
iron-scintillator sampling hadronic calorimeter, an iron absorber and
a set of 10 muon chambers. The detector was located within a magnetic
field of 0.4 T, perpendicular to the beam axis
for momentum determination. In fact, the magnetic dipole hosting the
NOMAD drift chambers was originally built for the UA1 experiment and
it is currently used by the T2K Collaboration for the 280~m near
detector~\cite{collaboration:2010hi}.

\begin{figure}[htbp]
\begin{center}
\rotatebox{0}{ \resizebox{0.8\textwidth}{!}{
\includegraphics{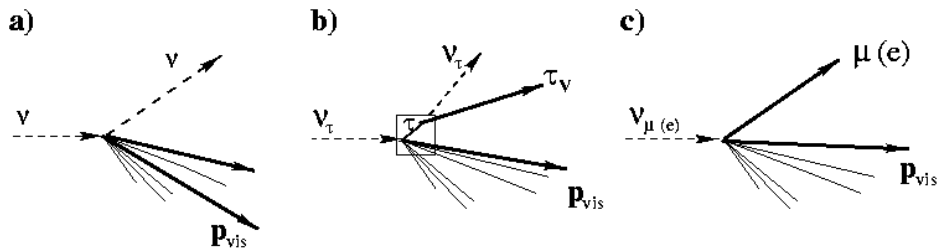}
}}\caption{\small Signal and background topologies in NOMAD: a) NC background; 
b) $\nu_\tau$ CC signal with subsequent $\tau$ decay; c) 
$\nu_{\mu}(\nu_{e})$ CC background. The square indicates 
the reconstructed ``primary" vertex for $\nu_\tau$ CC interactions. }
\label{fig:bkgnds}
\end{center}
\end{figure}

The NOMAD experiment based its search for $\nu_\tau$ on kinematical
criteria. From the kinematical point of view, $\nu_\tau$ CC events in
NOMAD are fully characterized by the decay products of the primary
$\tau$.  The spatial resolution of NOMAD did not allow the observation
of a secondary vertex from $\tau$ decay. The presence of visible
secondary $\tau$ decay products, ${\tau_{\rm V}}$, marks a difference
with respect to NC interactions, whereas the emission of one (two)
neutrino(s) in hadronic (leptonic) $\tau$ decays provides
discrimination against $\nu_\mu$ ($\nu_e$) CC interactions
(Fig.~\ref{fig:bkgnds}).  Hence, in $\nu_\tau$ CC events the
transverse component of the total visible momentum and the variables
describing the visible decay products have different absolute values
and different correlations with the remaining hadronic system, $H$,
than in $ \nu_\mu$ ($\nu_e$) CC and NC interactions.  The optimal
separation between signal and background is achieved when all the
degrees of freedom of the event kinematics (and their correlations)
are exploited.

The maximum $\nu_\tau$ signal allowed by limits from previous
experiments~\cite{Ushida:1986zn, McFarland:1995sr} was at least a
factor of 0.0025 times smaller than the main $\nu_\mu$ CC component and a
rejection power against backgrounds of $\mathcal{O}(10^5)$ was required
from the kinematic analysis.  Therefore, the $\nu_\tau$ appearance
search in NOMAD was a kinematic-based search for rare events within a
large background sample, as in Super-Kamiokande nowadays (see
Sec.\ref{sec:inclusive}). Similarly, in order to obtain reliable background estimates
the Collaboration developed methods to correct Monte Carlo (MC)
predictions with experimental data and defined appropriate control
samples to check such predictions.

%
\subsection{The CHORUS experiment}
\label{expe_chorus}

\begin{figure}[htbp]
\begin{center}
\rotatebox{0}{ \resizebox{0.8\textwidth}{!}{
\includegraphics{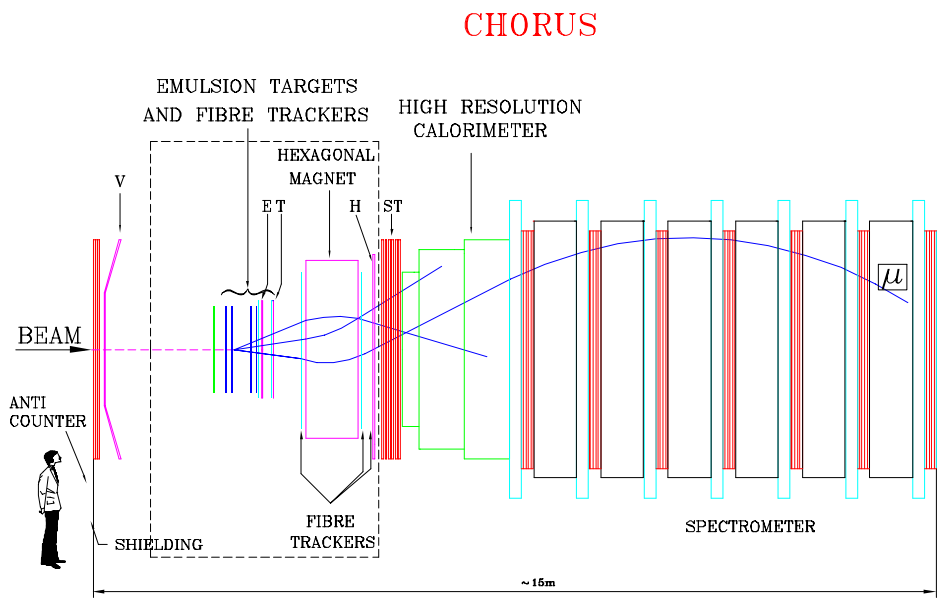}
}}\caption{\small  General layout of the CHORUS detector.}
\label{fig:detector}
\end{center}
\end{figure}

The approach followed by the CHORUS experiment was rather
different. Instead of relying on a kinematic analysis, a detector
based on nuclear emulsions with an ultra-high granularity
(1~$\mu$m) was employed.  A schematic picture of the CHORUS apparatus
is shown in Fig.~\ref{fig:detector}.

The {\it hybrid} setup was
made of an emulsion target (800 kg), a scintillating fibre tracker system,
trigger hodoscopes, a magnetic spectrometer, a lead-scintillator
calorimeter and a muon spectrometer.

The nuclear emulsions acted as the target and, simultaneously, as the
detector of the interaction vertex and of the $\tau$ lepton
decay~\cite{emul}. The emulsions were subdivided in $4$ stacks of $36$
plates, oriented perpendicularly to the beam and with a surface of
$1.44 {\times} 1.44~{\rm
\mbox{m}^2}$. Each plate was made of a $90~\mu$m transparent plastic
film with 350~$\mu$m emulsion sheets on both sides.  

The nuclear emulsion target was equip\-ped with a high resolution
tracker made of interface emulsions and scintillating fibre
planes.  Each stack was followed by three special interface emulsion
sheets: two Changeable Sheets (CS), close to the fibre trackers, and a
Special Sheet (SS), close to the emulsion stack. The sheets had a
plastic base of $800~\mu$m coated on both sides by $100~\mu$m emulsion
layers. Eight planes of target trackers of scintillating fibres
($500~\mu$m diameter)~\cite{fiber}, interleaved between the emulsion
stacks, measured the trajectories of the charged particles with a
precision of $150~\mu$m in position and 2~mrad in angle at the surface
of the CS. 

Downstream of the target region, a magnetic spectrometer was used to
reconstruct the momentum and sign of charged particles.  A hexagonal
air-core magnet~\cite{hex} produced a pulsed homogeneous field of
$0.12$~T. Field lines were parallel to the sides of the hexagon and
the magnetized region extended for a depth of $75$~cm in the direction
of the beam.  The tracking before and after the magnet was performed
by a high resolution detector made of scintillating fibres ($500~\mu$m
diameter) and complemented with a few planes of electronic detectors
(streamer tube chambers in the 1994, 1995 and at the beginning of 1996
run, honeycomb chambers~\cite{hcc} afterward).  The resulting momentum
resolution $\Delta p/p$ was $30\%$ at $5$~GeV.  \par In addition to
the detection elements described abo\-ve and in order to perform a
more precise kinematical analysis of the $\nu_\tau$ decay candidates,
the air-core hexagonal magnet region was equip\-ped with large area
emulsion trackers during the 1996 and 1997 runs.  \par A
$100~\mbox{ton} $ lead-scintillating fibre calorimeter~\cite{calo},
together with a lead-scintillator calorimeter, followed the magnetic
spectrometer and measured the energy and direction of electromagnetic
and hadronic showers, together with a lead-scintillator calorimeter.
\par A muon spectrometer made of magnetized iron disks interleaved
with plastic scintillators and tracking devices was located downstream
of the calorimeter. A momentum resolution of $19\%$ was achieved by
magnetic deflection for muons with momenta greater than $7$~GeV. At
lower momenta, the measurement of the range yielded a $6\%$
resolution.

\begin{figure}[htbp]
\begin{center}
\rotatebox{0}{ \resizebox{0.8\textwidth}{!}{
\includegraphics{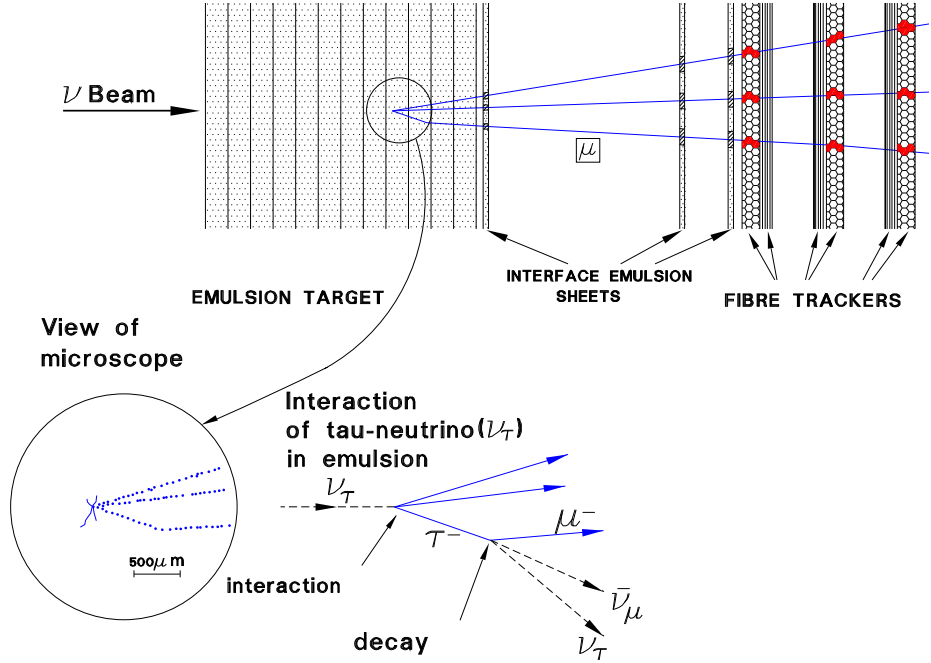}
}}\caption{\small $\nu_\tau$ detection principle exploited by the CHORUS experiment. A $\tau$ lepton produced in a $\nu_\tau$CC interaction produces a track of few hundred $\mu$m before its decay. The short $\tau$ track  and the decay product(s) are clearly visible thanks to the excellent spatial resolution of nuclear emulsions.}
\label{fig:targetzoom}
\end{center}
\end{figure}

CHORUS took advantage of the excellent spatial resolution (below 1
$\mu$m) of nuclear emulsions. Indeed, at the average energy of the
WANF neutrino beam the $\tau$ lepton produced in a $\nu_\tau$~CC
interaction travels about 1 mm before its decay. Both the parent track
($\tau$ lepton) and the decay product(s) can be seen in the nuclear
emulsion and the peculiar decay topology fully reconstructed as shown
in Fig.~\ref{fig:targetzoom}. The main advantage with respect to
NOMAD is the capability of identifying tau candidates on an
event-by-event basis, minimally relying on the kinematic analysis.

As discussed in detail in Ref.~\cite{Eskut:2007rn}, the main source of
background in the CHORUS experiment originated from the poor
efficiency in measuring the momentum and the charge of the decay
products. This was mainly due~\cite{Eskut:1997ar} to the limited
geometric efficiency of the spectrometer (85\%) and to the actual
performance of the trackers, which provided a charge discrimination
beyond $3\sigma$ only for momenta lower than 5~GeV.  Indeed, the
excellent sensitivity of the nuclear emulsions in detecting also
nuclear recoils (which is extremely important in rejecting hadron
reinteractions mimicking a decay topology) was partially compromised
by the low efficiency in performing the kinematical analysis of the
events. The background level achieved (normalized to charged-current
interactions) was $\approx10^{-3}$.

%
\subsection{CHORUS and NOMAD results}
\label{chorusnomadresults}

\begin{figure}[htp]
 \begin{center}
     \resizebox{1.0\textwidth}{!}{
	\includegraphics{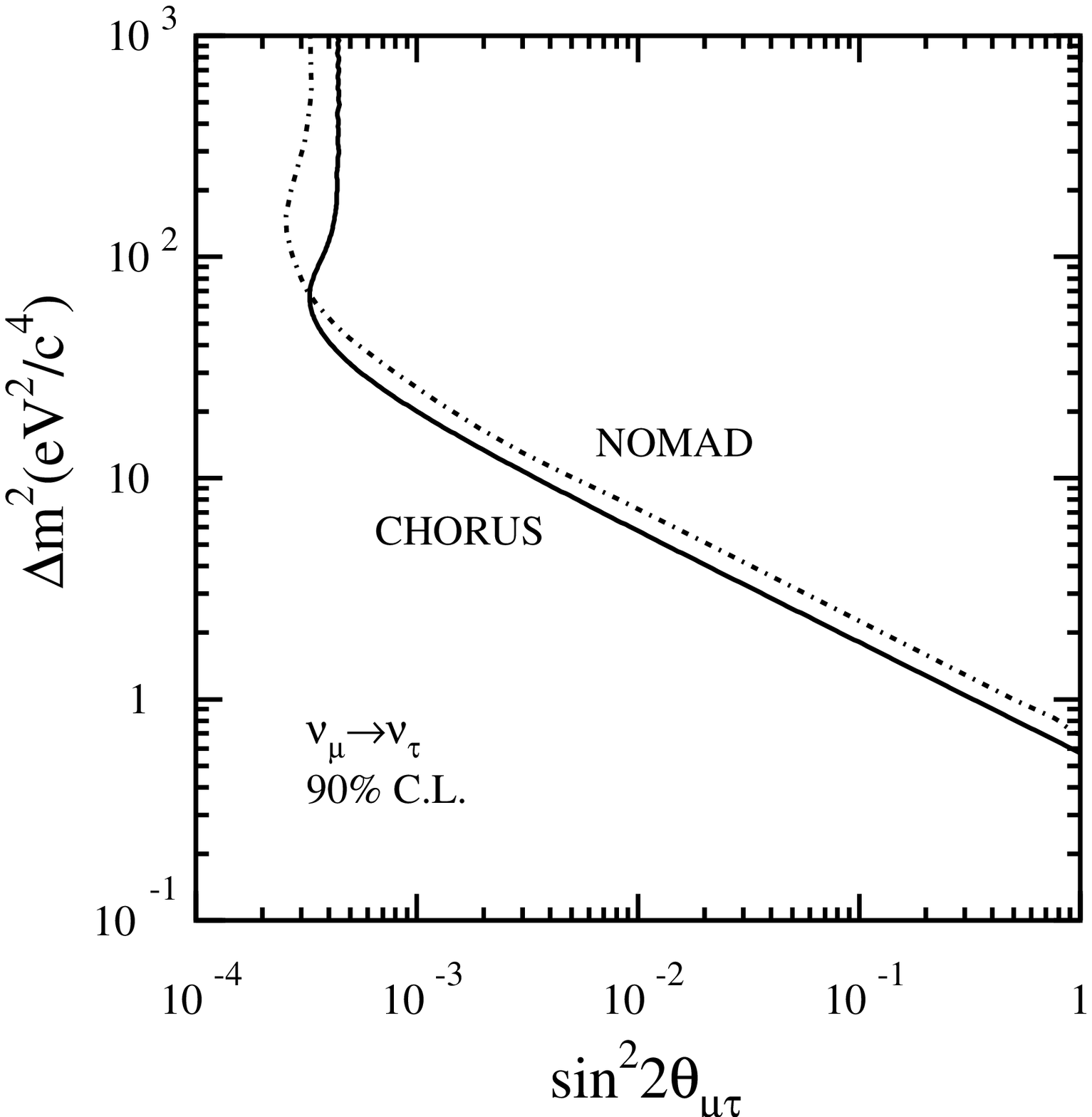} \includegraphics{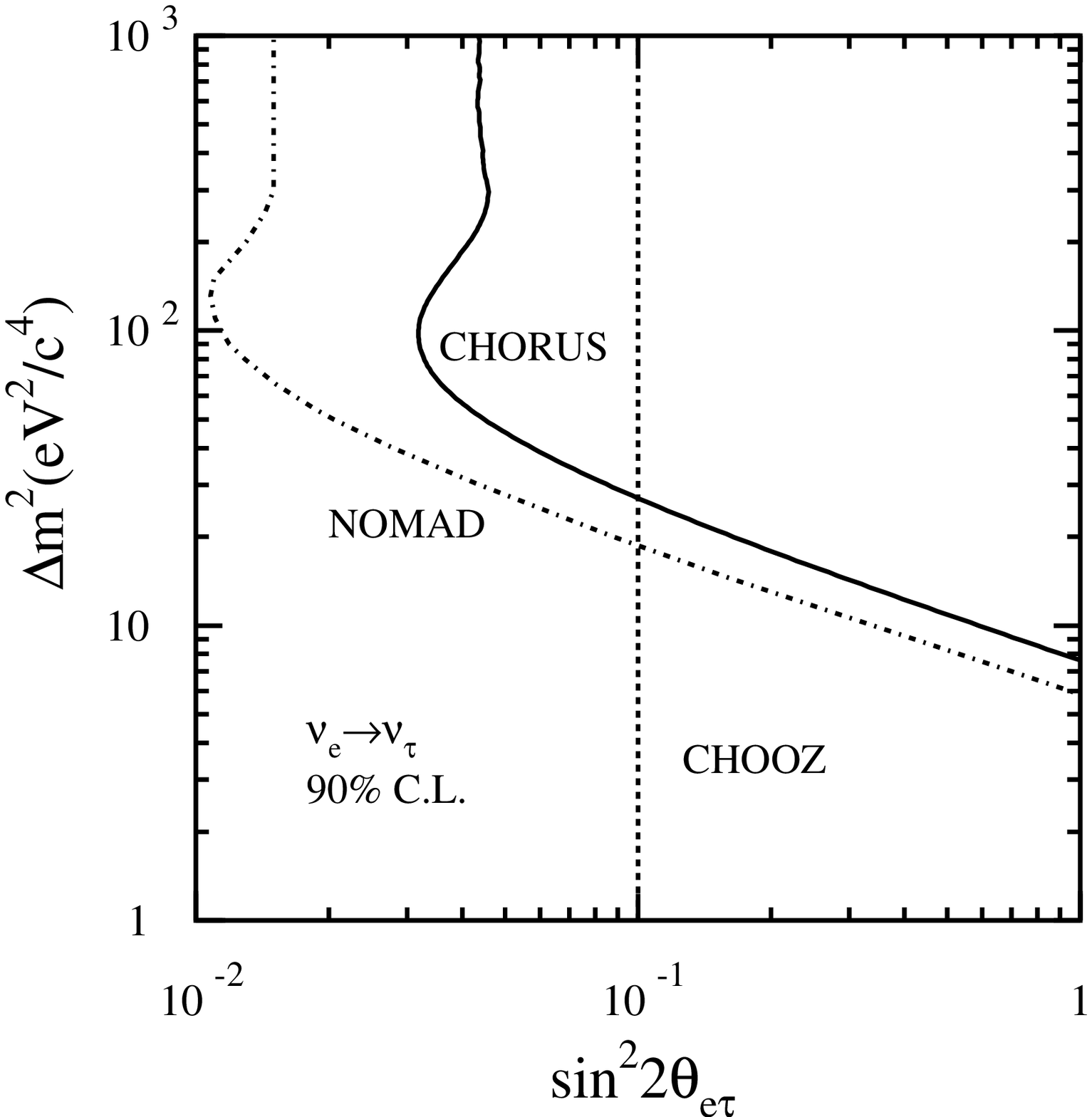}    }
   \caption{CHORUS (solid lines) and NOMAD (dashed lines) upper limits on $\nu_\mu \rightarrow \nu_\tau$ (left) and $\nu_e \rightarrow \nu_\tau$ 
(right) oscillation represented in an exclusion plot in the oscillation parameter plane.  The latter result is also compared with CHOOZ \cite{chooz}. }
   \label{fig:exclusionplot}
\end{center}
\end{figure}

Eventually, both 
CHORUS~\cite{Eskut:2007rn} and NOMAD~\cite{Astier:2001yj} had no evidence for
$\nu_\mu\rightarrow\nu_\tau$ and $\nu_e\rightarrow\nu_\tau$
oscillations.
The 90\%~C.L.\ upper limit on the appearance probability was
$$P (\nu_\mu \rightarrow \nu_\tau) < 1.6 \times 10^{-4} \, .$$
The corresponding limit on the mixing angle is quite stringent; it is
$\sin^2 2\theta_{\mu\tau} < 4 \times 10^{-4}$ for large $\Delta m^2$,
while the lower $\Delta m^2$ value excluded by the two experiments is
$\simeq 0.5~\mbox{eV}^2$ for maximal mixing (see
Fig.~\ref{fig:exclusionplot}). It is worth noting that at small mixing
angles NOMAD has a better sensitivity, while CHORUS is more sensitive
at low $\Delta m^2$. This difference is due to the different
experimental techniques employed by the two experiments. NOMAD has
been able to collect and analyze a much larger neutrino interaction
sample, while CHORUS has higher efficiencies at low neutrino energies.

The study of $\nu_e\rightarrow\nu_\tau$ oscillations was possible
thanks to the 0.9\% $\nu_{e}$ contamination of the SPS
neutrino beam.  Assuming that all observable $\nu_\tau$ would
originate from this contamination, the above result translates into a
limit on the $\nu_{e} \rightarrow \nu_\tau$ appearance
probability. The difference in energy between the $\numu$ ($\langle
E_{\numu} \rangle \sim 26$~GeV) and $\nu_e$ ($\langle E_{\nu_e}
\rangle \sim 42$~GeV) components leads to a different shape of the
exclusion plot in the oscillation parameter plane.
The corresponding 90\%~C.L.\ upper limit on the appearance probability is
$$P (\nu_e \rightarrow \nu_\tau) < 1 \times 10^{-2} \,.$$

The main reason of the large difference in the sensitivity between
CHORUS and NOMAD (see Fig.~\ref{fig:exclusionplot}) is due to the
harder $\nu_e$ spectrum with respect to the $\nu_\mu$ one and 
to the lower efficiency of CHORUS at high energies.

%
\section{Evidence from inclusive measurements}
\label{sec:inclusive}

Atmospheric neutrinos provided the first convincing evidence for
flavor transitions in 1998~\cite{sk_1998,atm}. In fact, $\numu$ and
$\nue$ produced by the interaction of primary cosmic rays with the
nuclei in the earth atmosphere are a powerful discovery tool: their
energy spans several order of magnitudes, the flux dropping as $E^
{-2.7}$, and oscillations can be probed from medium baselines -
$\mathcal O(10)$~km for neutrinos produced just above the detector -
up to a length comparable with the earth diameter for neutrinos
produced on the other side of the earth.  The characteristic
oscillation frequency due to the squared mass difference of the second
and third mass eigenstates (\dmtt) lies within the atmospheric
energy-baseline range and it is prominent in the multi-GeV region. It
manifests as a deficit of $\numu$ for zenith angles larger than
$\pi/2$ (up-going events). This depletion is absent in atmospheric
$\nue$ and, when combined with reactor data, excludes the occurrence
of a large \numunue conversion, which in turn demonstrates that
$\thetaot \ll \thetatt$. In 2003 this result has been confirmed with
artificial sources by the K2K experiment~\cite{K2K} and, more
recently, by MINOS~\cite{minos}. The measurement of MINOS, in
particular, provides the most precise measurement to date of $\dmtt$
($\dmtt = 2.32^{+0.12}_{-0.08} \times 10^{-3} \
\mathrm{eV}^2$~\cite{minos_adamson}). Similarly, atmospheric and
accelerator data~\cite{sterile} exclude sizable oscillation
probabilities into new types of neutrinos that are singlet under the
electroweak gauge group (``sterile neutrinos'') and point toward a
disappearance pattern that follows the sinusoidal law characteristic
of oscillations~\cite{sk_pattern}. We therefore expect the large
disappearance of $\numu$ in the multi-GeV range to be due to a large
\numunutau\ transition rate. Such conversion could be revealed in a
direct manner by the observation of atmospheric-induced tau leptons
from \nutau~CC events occurring in the detector.

Initial state atmospheric neutrinos are almost devoid of
\nutau. Tau neutrinos can be produced by the leptonic decay of
$D_s$ in $p+N$ interactions and the level of contamination does not
exceed $10^{-6}$~\cite{pasquali}. In addition, the rate of
$\nutau$~CC interaction is heavily suppressed by the large kinematic
threshold and the $E^{-2.7}$ damping factor in the $\numu$
spectrum. Assuming the current best fit value of $\dmtt$ and maximal
atmospheric mixing ($\thetatt = \pi/4$ and $\thetaot =0$), about 
1~\nutau\ CC event/year is expected in a detector with a 1~kton
mass~\cite{sk_app}. On top of this, the identification of the tau
lepton on an event-by-event basis is impossible with coarse-grained
detectors as the ones employed for the study of atmospheric neutrinos.
The tau lepton decays promptly (lifetime: 291$\pm$1~fs) in a variety of
final states that are briefly summarized in Table~\ref{tab:BRtau}. As
a consequence, unlike in $\nue$ and $\numu$ CC interactions, the final
state lepton in \nutau\ CC (the tau) is unobservable in the detector and can only be
identified by the topology or kinematics of its decay products. Still, the vast
majority of the decays are characterized by a one-prong topology,
i.e. one long-living charged particle (e, $\mu$ or $\pi^-$)
accompanied by the large missing energy that is carried by final state
neutrinos. This feature is exploited both by inclusive and exclusive
measurements (see Sec.~\ref{sec:exclusive}). Unfortunately, a
one-prong + missing energy topology can be easily mimicked by neutral-current (NC)
interactions, which represent the dominant background of any
inclusive analysis, while the signal-to-noise ratio is ultimately limited by
the granularity of the detector.

\begin{table}
\begin{center}
\begin{tabular}{c|c}
Decay Channel & Branching ratio (\%) \\
$\mu^- \anumu \nutau$ & $17.36 \pm 0.05$ \\
$e^- \anue \nutau$ & $17.85 \pm 0.05$ \\
$\pi^- \nutau$ & $10.91 \pm 0.07$ \\
$h^- \pi^0 \nutau $ & $25.94 \pm 0.09$ \\
$h^- \ge 2 \pi^0 \nutau $ & $10.85 \pm 0.12$ \\
$h^- h^+ h^- \ge 0 \ \mathrm{neutrals}\ \nutau$  & $14.56 \pm 0.08$ \\
\end{tabular}
\label{tab:BRtau}
\caption{Main decay channels for the tau lepton observed in direct appearance searches~\cite{PDG}.
$h^\pm$ stands for $\pi^\pm$ or $K^\pm$.}
\end{center}
\end{table}

At the atmospheric scale, evidence for tau appearance has been gained
by the SuperKamiokande experiment using water Cerenkov techniques.
The huge fiducial mass of the detector (22.5 kton) combined with a decade long
exposure compensate the poor signal-to-noise ratio. Larger purities
could be obtained by fine-grained detector similar to NOMAD (see
Sec.~\ref{sec:early}) or by homogeneous liquid-Argon
TPC's~\cite{lar}. Still, the size of these detectors cannot compete
with SuperKamiokande: the weight of the largest liquid-Ar detector
presently under operation is just 600~tons~\cite{600t}. Multi-kton
liquid-Ar detectors are, however, considered a viable option to
overcome the limitations of water Cherenkov detectors in future atmospheric
and long-baseline accelerator experiments.  In particular, the opportunities offered
by this technology at the 10~kton scale for the inclusive measurement
of tau appearance with atmospheric neutrinos has been recently discussed
in~\cite{Conrad:2010mh}.

The SuperKamiokande Collaboration has published its first analysis on
tau appearance in 2006~\cite{sk_app} after an exposure of 1489.2 days
(``Super-K I data taking''). The detector consists of two concentric,
optically separated regions filled with radiopure water and read-out
by large photomultipliers (PMT): the inner region is employed for
vertex location and it is equipped with about 11000 20-inch PMT's
(``inner detector''); the outer - readout by sparser 8-inch PMT's -
is used to veto cosmic-ray background, shield neutrons and $\gamma$'s
from the surrounding rock (``outer detector'') and identify partially
contained events. The overall detector mass is 50~kton but the
fiducial volume includes only reconstructed vertices with minimum
distance from the walls of the inner detector of 2~m. It corresponds
to an effective mass of 22.5~kton. SuperKamiokande identifies charged
particles by the corresponding Cherenkov rings in water. In
particular, the \nutau\ analysis aims at selecting an enriched sample
of hadronically decayed tau leptons. Semileptonic tau decays as $\tau
\rightarrow \mu \numu \nutau $ or $\tau \rightarrow e \nue \nutau $
are not employed due to the overwhelming background of \numu\ and
\nue\ CC interactions. Minimum ionizing particles (mainly muons and
charged pions) produce sharp ring edges with variable openings, while
(ultrarelativistic) electrons and converted photons generate diffused
ring patterns with a fixed opening angle of 42$^\circ$. $\nutau$~CC
interactions occur at $E>3.5$~GeV and, therefore, are mostly dominated
by deep-inelastic scattering. Except for the $\tau \rightarrow \mu
\numu \nutau $ decay channel, most $\nutau$ events show a multi-ring
topology without a leading $\mu$-like (sharp) ring (``e-like
sample''). SuperKamiokande has, thus, performed its inclusive analysis
in a subsample of events having the vertex located inside the fiducial
volume, a visible energy greater than 1.33~GeV and the most energetic
ring clearly identified as e-like. This subsample has a signal
($\nutau$~CC events) to noise ratio of 3\% for maximal
mixing and $\dmtt=2.4 \times 10^{-3}$~eV$^2$. Further
enrichment can be obtained considering kinematic variables that
enhance the difference between tau-like decay topologies and NC or
$\nue$~CC events. In particular, five variables have been considered
in \cite{sk_app}: the visible energy, the maximum distance between the
primary interaction and electron vertices from pion and then muon
decay, the number of rings, the sphericity in the laboratory frame and
the clustered sphericity in the center-of-mass frame. Shape
information from these variables are combined in a likelihood or,
equivalently, in a neutral-network output and ``tau-like'' events are
defined as candidates with a likelihood (NN output) greater than 0
(0.5). The tau-like final subsample has a signal-to-noise ratio of
about 5\%. Although this analysis is in principle strongly dependent
on detector simulation, the sample of down-going events provides a
unique tool for MC validation. Down-going events are generated by
neutrinos that were produced just above the detector, at an average
height of 15~km from sea-level if they originate from the decay in
flight of $\pi$, 13~km if they arise from muon decays.  At these
baselines the \numunutau\ probability is negligibly small and,
therefore, they represent a pure sample of unoscillated
neutrinos. Evidence of tau appearance can be drawn by the binned fit
of the zenith angle ($\theta$) distribution of tau-like events.  The
fit is done assuming an arbitrary overall normalization
($\alpha,\beta$) both for signal ($N^{tau}_i$) and background
($N^{bkg}_i$), i.e. minimizing
\be
\chi^2 = \sum_{i=1}^n \frac{ (N^{obs}_i -\alpha N_i^{tau}-\beta N_i^{bkg})^2 }{\sigma_i^2}
\ee
$\sigma_i$ being the statistical error for the $i$-th bin. A large
$\mathrm{Min}(\chi^2)$ for the null hypothesis ($N^{tau}_i=0$ for any
i) is an indication of regions rich of oscillated $\nutau$. In
fact (see Fig.~3 of Ref.~\cite{sk_app}), the zenith distribution shows
a rather clear excess of tau-like events in the up-going region,
i.e. in the region where oscillations are expected to be large. Thanks
to the good knowledge of the leading oscillation parameters \dmtt\ and
$\thetatt$ and to the up/down comparison of the rates, the
contributions to the systematic error that in principle should
dominate the measurement (the $\sin^2 2\theta_{23}$ factor and the
flux overall normalization) mostly cancel out. In fact, systematics
are dominated by unknown size of the \thetaot
angle and by the uncertainties on the $\nutau$ cross-section.

A non-zero value of the $\thetaot$ mixing angle between the first and
third family causes a \numunue conversion that enriches the e-like
sample. The enhancement cannot be corrected by the up/down comparison
because it is driven by the same phase $\Delta_{23}$ as for the
leading $\numunutau$ oscillations and, therefore, builds up only for
large baselines (up-going).  Note also that the systematic shift due
to \thetaot is asymmetric since, for any value of the angle, it always
causes an apparent increase of the statistics of up-going
events. SuperKamiokande estimated this effect to be smaller than
21\%. The estimate relies on the present best limit on \thetaot from
the CHOOZ experiment ($\sin^2 \theta_{13} < 0.027$ at 90\%
CL~\cite{chooz,Schwetz:2008er}). Since the measurements of \thetaot by
reactor or long-baseline experiments are uncorrelated with the
atmospheric measurement, a significant improvement of this systematics
is expected in the next few years, taking advantage of the results
from Double-CHOOZ~\cite{doublechooz}, RENO~\cite{reno},
T2K~\cite{t2k}, Daya Bay~\cite{dayabay} and NO$\nu$A~\cite{nova}. The
effect being proportional to $\sin^2 2\thetaot$, we expect such
systematics to drop at the level of a few percent in less than 5
years. On the contrary, no major improvements can be anticipated from
the other dominant contribution, i.e. the knowledge of the $\nutau$
cross-section. Although this number is immaterial when testing against
the no-appearance (null) hypothesis (i.e. the hypothesis corresponding
to no evidence for taus in the enriched sample), it is of relevance in
order to compare the data to the expected rate within the standard
three-family oscillation framework. SuperKamiokande is
particularly sensitive to the uncertainty in the cross-section because
the oscillated tau-events are mostly at low energy, i.e. in the
proximity of the sharp rise of the cross-section just beyond the
kinematic threshold. This effect is due to the $E^{-2.7}$ cutoff of
the unoscillated $\numu$ spectrum. SuperKamiokande estimated this
contribution to be smaller than 25\% by comparison among different
theoretical models~\cite{xsect}.

The 2006 (SuperK-I) analysis provided evidence for tau appearance at
the 2.4 sigma level, the best fit of the signal in the tau-like sample
being $138 \pm 48\mathrm{(stat)} ^{+15}_{-32}\mathrm{(sys)}$. A priori, the
expected sensitivity was about 2~sigma and a larger rejection
power against the null hypothesis has been reached thanks to the
larger number of observed events.

In 2001, during the refill after a shutdown aimed at replacing dead
photomultipliers (PMT), an accident occurred in the SuperKamiokande
detector, which caused the loss of about 60\% of the
photodetectors. The remaining 20-inch PMT were redeployed in the inner
detector (ID) while the PMT of the outer veto were fully rebuilt. In
this configuration, which is clearly not optimal for inclusive tau
search due to the reduced ID coverage (47\% of the original one)
SuperKamiokande took data until 2005 (``SuperK-II''). Still, the
atmospheric results obtained during SuperK-II are in good
agreement with previous results both in the
disappearance~\cite{Itow:2008zza} and in the
appearance~\cite{kato_thesis} analyses. The repair of SuperKamiokande
was completed in July 2006 and, since then, full coverage at the inner
detector has been restored (``SuperK-III data taking'').  The data
taking of this third phase was completed in 2008 and the preliminary
SuperK-I,II,III combined data have been presented in December
2010~\cite{nnn2010}. SuperK-III adds 518 days of statistics at nominal
coverage; moreover, the inclusive tau appearance analysis has
been improved. It now employs a 2-D unbinned likelihood fit that uses
the complete distribution of the neural-network output instead of a
sharp cut ($>0.5$) to distinguish between background-like and tau-like
topologies. It also makes use of a modified set of inclusive variables
with higher sensitivity. The overall expected sensitivity of the new
analysis computed from simulation and assuming nominal cross-sections
is significantly better than the one of 2006 (2.6$\sigma$ versus
2$\sigma$). On top of this, a larger enhancement has been observed
in the up-going sample with respect to the down-going reference
data. Such enhancement excludes the null hypothesis (no tau
oscillation) at the 3.8$\sigma$ level. If these preliminary results
are confirmed, we can safely expect that inclusive measurements will
dominate the experimental evidence for \numunutau\ transitions in the
next few years, before the final results from CNGS.

\section{Evidence from exclusive measurements}
\label{sec:exclusive}
Inclusive analyses try to distill a tau-enriched sample in
the bulk of \numu\ and \nue\ interactions and take advantage of the
large statistics and of the peculiar kinematics of tau
decays. Exclusive measurements are even more ambitious since they aim
at observing the appearance of tau leptons on an event-by-event basis.
It necessarily requires a detector with very high spatial resolution,
such as to observe the decay in flight of the tau and, at the same time, a
high intensity source with an energy well exceeding the kinematic
threshold for tau production. The only facility that is able to
fulfill simultaneously these requirements is the CNGS facility in
Europe. The CNGS beam~\cite{CNGS} is a pure \numu\ beam with a mean
energy of 17~GeV produced at CERN and pointing to the Gran Sasso
Laboratories of INFN in Italy (LNGS), 730~km away from the source. The
intrinsic \nutau\ contamination, mainly originating from the decay of
$D_s$, is negligible ($<10^{-6}$). The beam is also contaminated at the
0.8\% by \nue, resulting from the decay in flight of muons along the
decay tunnel and from $K_{e3}$ decays. Since the tau lepton is identified
from its decay topology, background from prompt \nue is irrelevant.

The observation of tau leptons at the far detector will thus prove
unambiguously that the $\nu_\mu\rightarrow\nu_\tau$ oscillation is the
dominant transition channel at the atmospheric scale.  This is the
main goal of the OPERA experiment \cite{operaold, proposal1,
 proposal2}, which has been built from 2004 to 2008~\cite{operafirst} in the
Hall C of LNGS as a far detector for CNGS.

In OPERA, the $\nu_\tau$ appearance signal is detected through the
measurement of the decay daughter particles of the $\tau$ lepton
produced in CC $\nu_\tau$ interactions. Since the short-lived $\tau$
particle has an average decay length of about 1 mm at the CNGS beam
energy, a micrometric detection resolution is needed.  In OPERA,
neutrinos interact in a large mass target made of lead plates
interspaced with nuclear emulsion films acting as high-accuracy
tracking devices. This kind of detector is historically called an
Emulsion Cloud Chamber (ECC) and it has been successfully applied by
the DONUT experiment to perform the first direct observation of
$\nutau$ charged-current interactions in a \nutau\ enriched beam at
Fermilab~\cite{donut}.

\begin{figure}
\begin{center}
 \includegraphics[width=15cm]{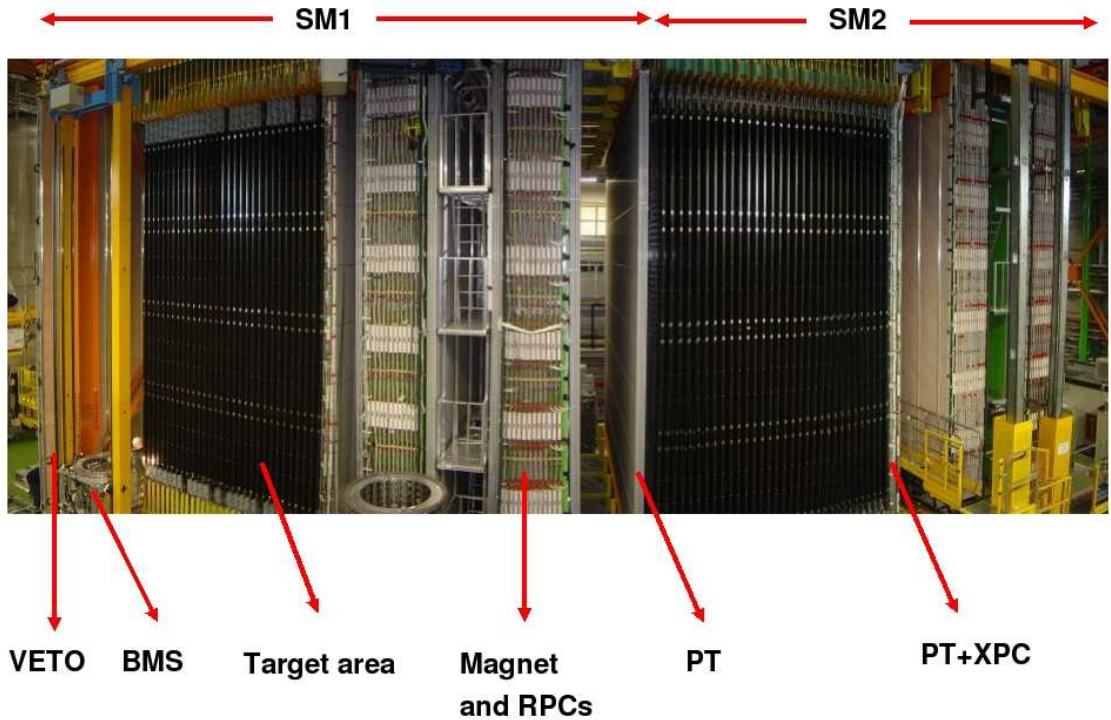}
  \caption{View of the OPERA detector. The upper horizontal lines
    indicate the position of the two identical supermodules (SM1 and
    SM2). The ``target area'' is made of walls filled with ECC bricks
    interleaved with planes of plastic scintillators. Arrows show the
    position of the VETO planes, the drift tubes (PT) pulled alongside
    the XPC, the magnets and the Resistive Plate Chambers (RPC) installed
    between the magnet iron slabs. The XPC are RPC planes with redout
    strips inclined by 45$^\circ$ with respect to the horizontal. The
    Brick Manipulator System (BMS) is also visible. } \label{detector}
\end{center}
\end{figure}

OPERA is a hybrid detector~\cite{operadetector} (see
Fig.~\ref{detector}) made of a veto plane followed by two identical
Super Modules (SM). Each SM consists of a target section of about 625
tons made of 75000 emulsion/lead ECC modules, or ''bricks'', of a
scintillator Target Tracker detector (TT) to trigger and
localize neutrino interactions within the target, and of a muon
spectrometer. A target brick consists of 56 lead plates of 1 mm
thickness interleaved with 57 emulsion films and with a mass of 8.3 kg. Their
thickness along the beam direction corresponds to about 10 radiation
lengths. In order to reduce the emulsion scanning load, Changeable
Sheet (CS) films have been used. They consist of tightly
packed doublets of emulsion films glued to the downstream face of each
brick. 

Charged particles from a neutrino interaction in a brick cross the CS
and produce signals in the TT that allow the corresponding brick to be
identified and extracted by an automated system. The hit patter in the
TT provides information on the bricks where the neutrino interaction
has occurred. The brick with the largest probability to contain the
vertex is, hence, extracted and its CS is detached, developed and
scanned. If tracks are found in the CS matching the
expectation from TT, the brick is developed and the tracks are traced
back up to the interaction vertex. Otherwise the procedure is repeated
for the second most probable brick.

 Large ancillary
facilities are used to bring bricks from the target up to the
automatic scanning microscopes at LNGS and various laboratories in
Europe and Japan \cite{ESS,SUTS}. Extensive information on the OPERA
detector and its support facilities are given in
\cite{operadetector,facilities}.

\begin{figure}
\begin{center}
 \includegraphics[width=15cm]{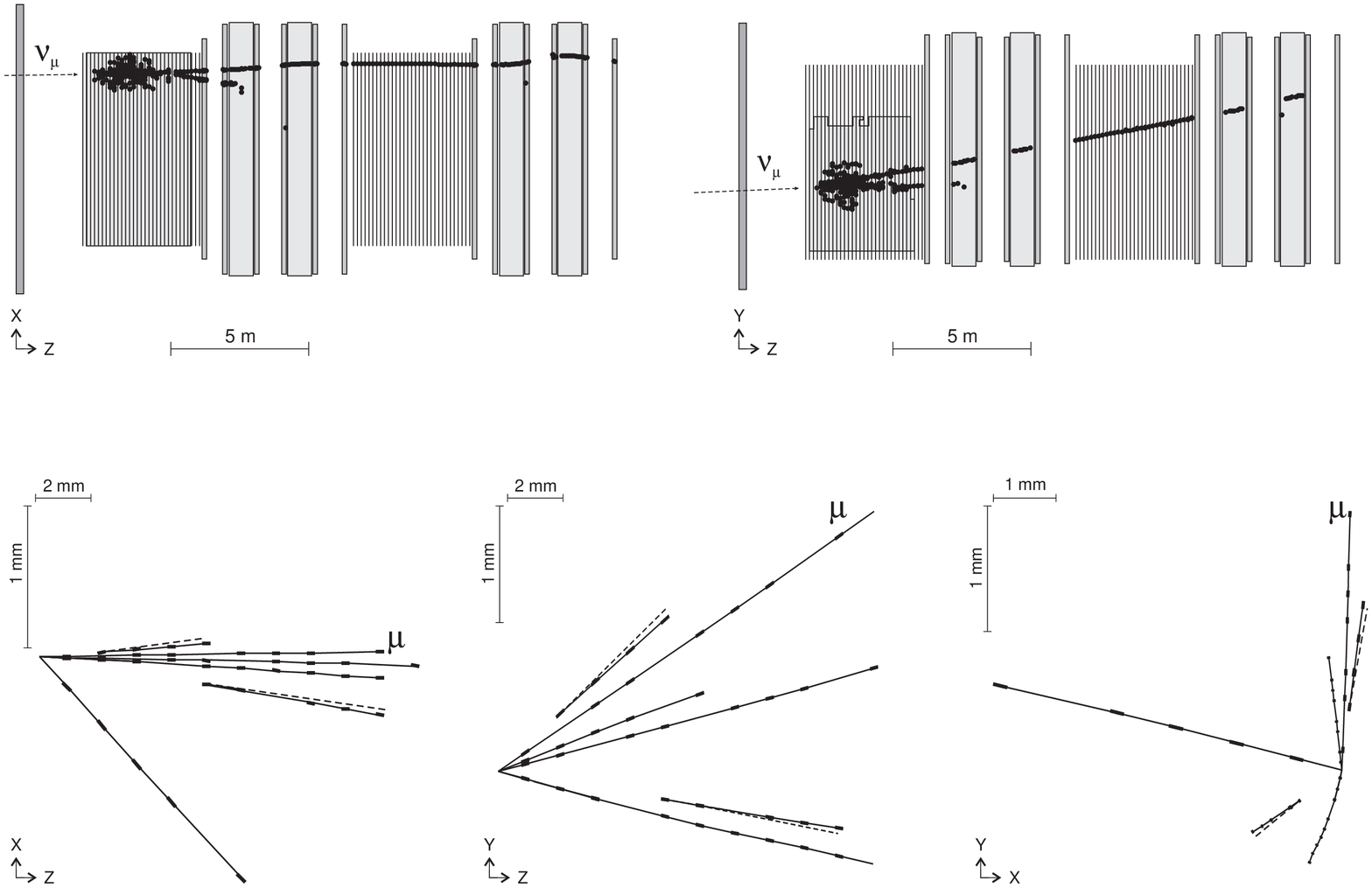}
  \caption{Top panels: on line display of an event seen by the OPERA
    electronic detectors (side and top views): a $\nu_\mu$ interacts
    in one of the first bricks of the first supermodule (SM) yielding
    hadrons and a muon which is detected in both SMs and whose
    momentum is measured by the magnets of the two SMs. Bottom
    panels: the vertex of the same event observed in the emulsion
    films (side, top and front views). Note the two
    $\gamma\rightarrow e^{+}e^{-}$ vertices: the opening angle
    between them is about 300 mrad. By measuring the energy of the
    $\gamma$'s one obtains a reconstructed invariant mass of
    $110\pm30$ MeV, consistent with the $\pi^{0}$
    mass.} \label{babyopera}
\end{center}
\end{figure}

A reconstructed CC event is shown in the bottom panels of
Fig.~\ref{babyopera}. In this case the dimensions of the event views are
of the order of a few millimeters, to be compared with the $\sim10$ m
scale of the whole event reconstructed with the electronic detectors
(top panels of Fig.~\ref{babyopera}).

As mentioned before, the CNGS beam is optimized for the observation of
$\nu_\tau$ CC interactions. The average neutrino energy is $\sim$17
GeV. The $\bar{\nu}_\mu$~CC contamination is 2.1\%; the $\nu_e$ and
$\bar{\nu}_e$ contaminations are less than 1\% and, as noted
above, the number of prompt $\nu_\tau$ is negligible.  With a total
CNGS beam intensity of $22.5\times 10^{19}$ protons on target
(p.o.t.), about 24300 neutrino events would be collected.

The $\tau$ decay channels investigated by OPERA cover all the decay
modes. Indeed, the e, $\mu$, single-prong (lines 3-5 of
Table~\ref{tab:BRtau}) and multi-prong (line 6 of
Table~\ref{tab:BRtau}) decays are measured. They are classified in 2
categories: ``long'' and ``short'' decays. Short decays correspond to
the cases when the $\tau$ decays in the same lead plate in which the
neutrino interaction occurred. The $\tau$ candidates are selected on
the basis of the impact parameter of the $\tau$ daughter track with
respect to the interaction vertex (IP $>$ 5-20 $\mu$m). In the long
decay category the $\tau$ does not decay in the same lead plate and
its track can be reconstructed in one film. The $\tau$ candidate
events are selected either on the basis of the existence of a kink
angle between the $\tau$ and the daughter tracks ($\theta_{kink}>$ 20
mrad) or on the presence of a secondary multi-prong vertex along the
$\tau$ track.

In order to improve the signal to background ratio, a kinematical
analysis is applied to $\tau$ candidates selected on the basis of the
topological criteria discussed above. Since for short decay candidates
the main background comes from charm production, a lower cut at
$2~\mbox{GeV}$ on the invariant mass of the hadronic system is
imposed. This cut reduces the background by more than a factor 1000,
while retaining about 15\% of the signal. For long decay candidates it
is worthwhile to consider leptonic, single-prong and multi-prong
decays separately.  For leptonic decays soft cuts on the daughter
momentum, a lower one to minimize the effect of the particle
misidentification ($p > 1$~GeV) and an upper one ($p < 15$~GeV) to
suppress the beam related background (\nue\ from the beam and the high
energy \numu\ tail of CNGS), and a soft cut on the measured transverse
momentum ($p_T$) at the decay vertex are enough to reduced a
background at a reasonable level. The applied cut at the decay vertex
is $p_T>100$~MeV and $p_T>250$~MeV for the electronic and muonic decay
channel, respectively.

For the single-prong decay the kinematical analysis is slightly more
complicated. The main background for this channel originates from the
reinteraction of primary hadrons without any visible recoil at the
reinteraction vertex. In order to keep the background for this channel
as low as possible kinematical cuts are applied both at the decay and
at the primary vertex. The kinematical analysis is qualitatively
similar to that of the electronic and muonic channels. However, the
cut applied on the $p_T$ is harder ($p_T>300$~MeV if a $\gamma$ is
attached to the decay vertex, $p_T>600$~MeV if not) and the daughter
particle is required to have a momentum larger than 2~GeV. The
kinematical analysis at the primary vertex uses the variables
$p_T^{miss}$, defined as the missing transverse momentum at the
primary vertex, and $\phi$, which is the angle in the transverse plane
between the parent track and the shower direction. Due to the
unobserved outgoing neutrino, $p_T^{miss}$ is expected to be large in
NC interactions. Conversely, it is expected to be small in CC
interactions. For $\tau$ candidates the measured $p_T^{miss}$ is
required to be lower than 1~GeV. The $\phi$ angle is expected to peak
at $\pi$, because the $\tau$ and the hadronic shower are back-to-back
in the transverse plane. On the contrary, the hadron mimicking a
$\tau\rightarrow h$ decay is produced inside the hadronic shower in NC
interactions. Therefore, $\phi$ peaks near 0 and for $\tau$ candidates
the $\phi$ angle is required to be larger than $\pi/2$.

For the multi-prong decay channel the main background is given by
multi-prong decay of charmed particles. The hadronic reinteraction
background is not a major issue. Indeed, the probability for a hadron
to undergo an interaction with multi-prong is much smaller (1-2 order
of magnitudes) than for single-prong interactions. The signal to
background ratio is enhanced by performing a kinematical analysis
mainly based on the following variables: $p_T^{miss}$, defined as the
missing transverse momentum at the primary vertex, the invariant mass
of the hadronic system, the total energy of the event.

\begin{figure}
\begin{center}
 \includegraphics[width=7.5cm]{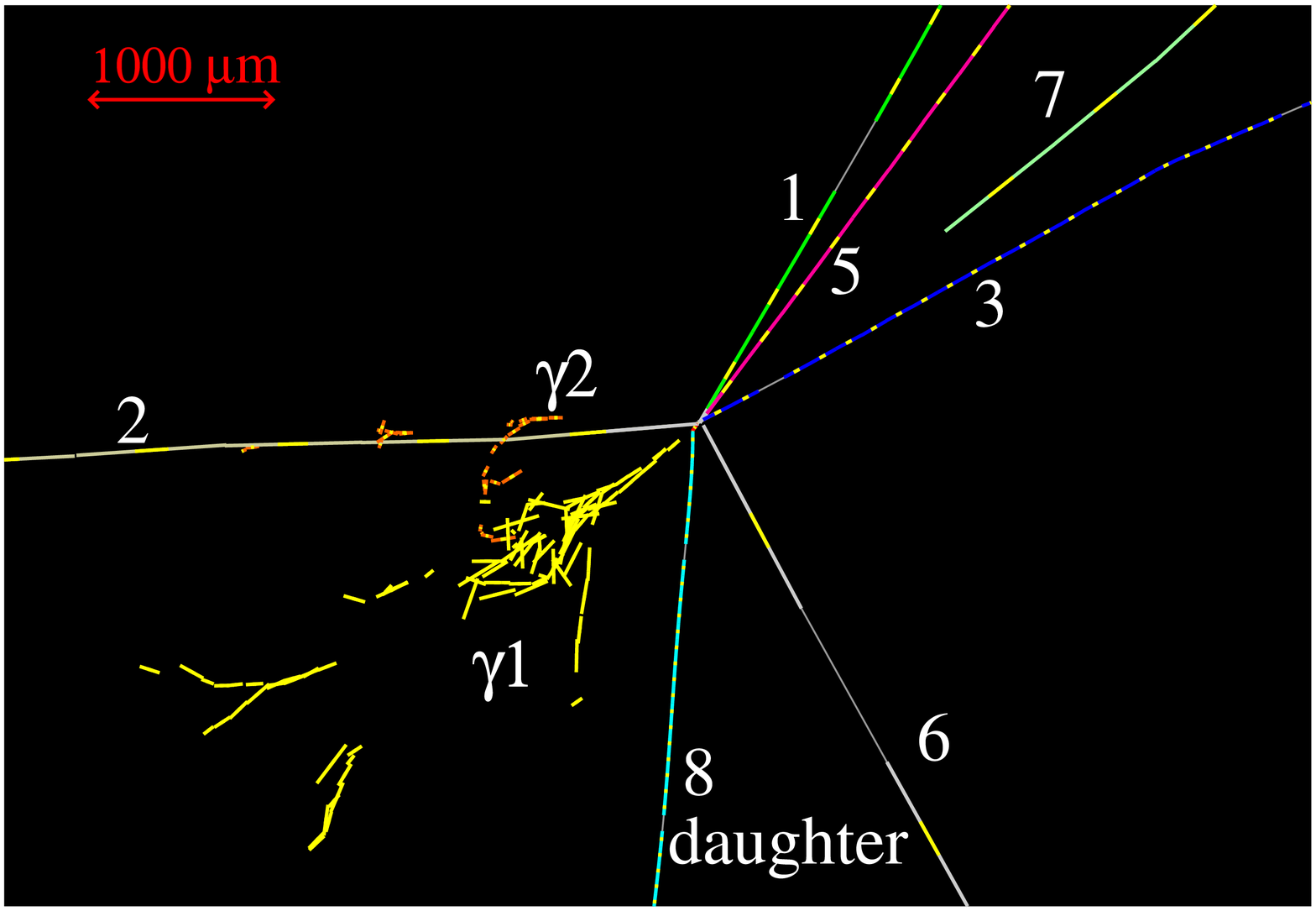}
   \includegraphics[width=7.5cm]{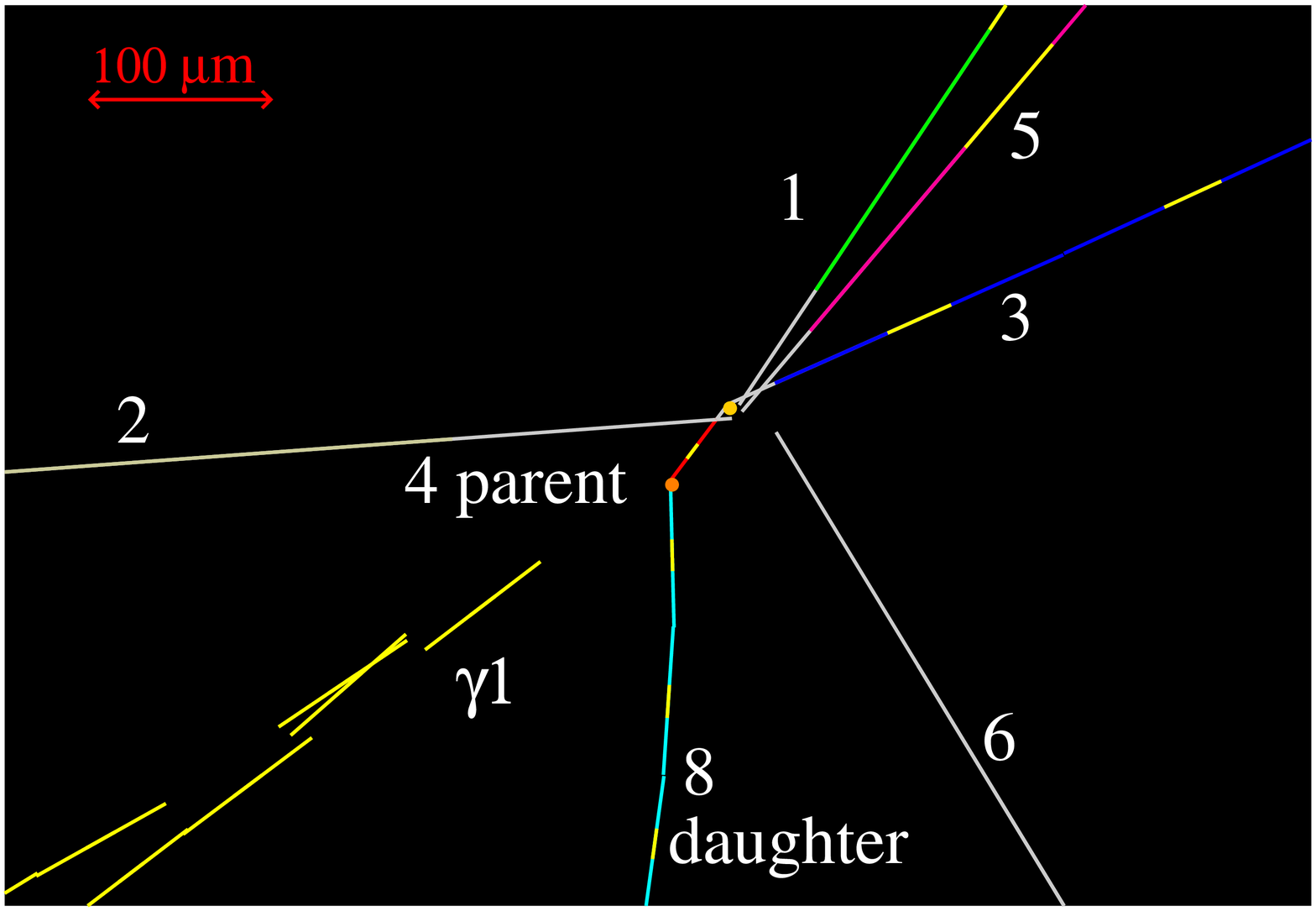}\\
     \includegraphics[width=15cm]{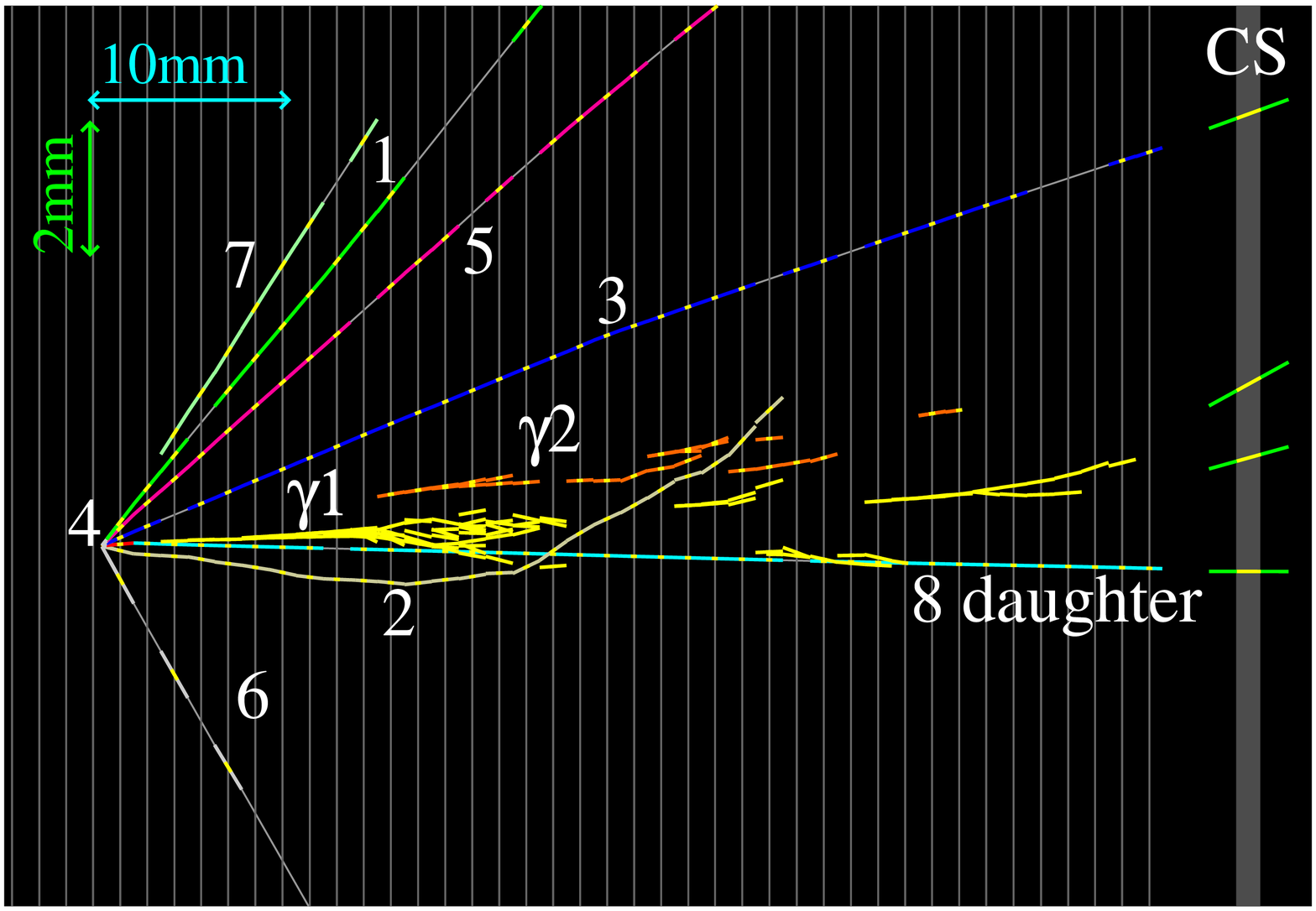}
  \caption{Display of the $\tau^-$ candidate event. Top left: view transverse to the neutrino direction. Top right: same view zoomed on the vertices. Bottom: longitudinal view.} \label{event}
\end{center}
\end{figure}

In Ref~\cite{Agafonova:2010dc} the OPERA Collaboration reported the
observation of a first candidate $\nu_\tau$ CC interaction in the
detector. The primary neutrino interaction consists of 7 tracks of
which one exhibits a visible kink. Two electromagnetic showers due to
$\gamma$-rays have been located; they are clearly associated with the
event and were produced at the decay vertex. Fig.~\ref{event} shows a
display of the event, which was identified in a sample corresponding
to $1.89\times10^{19}$ p.o.t. in the CNGS $\nu_\mu$ beam. The total
transverse momentum $P_T$ of the daughter particles with respect to
the parent track is ($0.47^{+0.24}_{-0.12}$) GeV, above the lower
selection cut-off at 0.3 GeV. The missing transverse momentum
$P_T^{miss}$ at the primary vertex is ($0.57^{+0.32}_{-0.17}$)
GeV. This is lower than the upper selection cut-off at 1 GeV. The
angle $\Phi$ between the parent track and the rest of the hadronic
shower in the transverse plane is equal to ($3.01\pm0.03$) rad,
largely above the lower selection cut-off fixed at $\pi$/2. The
invariant mass of $\gamma$-rays is ($120\pm20(stat.)\pm35(syst.)$)
MeV$^2$, supporting the hypothesis that they originate from a $\pi^0$
decay. Similarly the invariant mass of the charged decay product
assumed to be a $\pi^-$ and of the two $\gamma$-rays is
($640^{+125}_{-80}(stat.)^{+100}_{-90}(syst.)$) MeV, which is
compatible with the $\rho(770)$ mass. The branching ratio of the decay
mode $\tau\rightarrow\rho^-\nu_\tau$ is about 25\%.  The observation
of one possible tau candidate in the decay channel
$h^-(\pi^0)\nu_\tau$ has a significance of 2.36$\sigma$ of not being a
background fluctuation from a background of $0.018\pm0.007$. If one
considers all decay modes included in the search, corresponding to
$0.54 \pm 0.13$ expected taus, the significance of the observation
becomes 2.01$\sigma$ from the total predicted background of
$0.045\pm0.023$.

%
\section{Future experiments}
\label{sec:future}

\subsection{Tau appearance in the standard oscillation scenario}
\label{sec:future_standard}

The role of tau appearance for a direct proof of oscillation is
unique. In the standard three-family framework, precision measurements
of the leading oscillation parameters $\theta_{12}$, \thetatt, \dmot\
and \dmtt , can be done more easily studying the disappearance of
solar+reactor \nue and atmospheric+accelerator \numu. Similarly, the
unknown angle \thetaot and the CP violating phase $\delta$ will likely
be addressed studying subdominant \numunue (or \nuenumu) oscillations
at the atmospheric scale. This is the reason why there are no
facilities that have been pursued since 1998 and that are specifically
tuned for precision measurements of \numunutau\ transitions. In fact,
among the many setups proposed to study CP violation in the leptonic
sector, just a few~\cite{nufact,veryhigheBB} work beyond the kinematic
threshold for tau production. The most prominent are the Neutrino
Factories~\cite{nufact}, where neutrinos are produced by the
decay-in-flight of muons or antimuons. (Anti)muons produce final-state
neutrinos from the decay $\mu^+ \rightarrow e^+ \anumu \nue$. The
neutrino factories allow for the study of the transition $\nue
\rightarrow \numu$ if one identifies the signal of ``wrong-sign''
muons coming from \numu~CC events in the bulk of ``right-sign'' muons
originating from $\anumu$~CC interactions. As a consequence, the most
natural far detector for the neutrino factory is a high-density
magnetized calorimeter~\cite{MIND} with outstanding charge
reconstruction capabilities. The need of charge identification
efficiencies well above 99.9\% requires a strong muon energy cut to
filter punch-through pions or pions that have decayed in flight.  In
the classic high energy Neutrino Factory
configuration~\cite{ISS_phys,ISS_acc}, 50~GeV muons are accelerated
and stacked in a decay ring, producing $\nue$ of $\sim 30$~GeV
energy. In order to achieve charge misidentification of
$\mathcal{O}(10^{-4})$, a tight muon energy cut is applied so that the
detector efficiency drops to zero at energies below 10~GeV. For
typical baselines of 2000~km, it means that the peak of the
oscillation (neutrinos where $\dmtt L/4E \simeq \pi/2$) remains
completely unobserved in the Neutrino Factories, as well as in
CNGS. Such tight cut is the origin of one of the main drawback of the
Neutrino Factories: unlike more traditional setups that study \numunue
transitions at the oscillation peak (``Superbeams''), a strong
parameter degeneracy appears once we go from the measurement of the
\numunue and $\anue \rightarrow \anumu$ probabilities to the
determination of the $\delta$ and \thetaot\ angles (``intrinsic
degeneracy''~\cite{intrinsic_deg}). The degeneracy is particularly
severe for $\sin^2 2\thetaot \simeq 10^{-3}$, i.e. in a region where
the performance of the Neutrino Factory should be unbeatable with
respect to Superbeams~\cite{ISS_phys} or
Betabeams~\cite{betabeam}. Tau appearance can help to overcome this
drawback, at the price of a dedicated detector located at a shorter
baseline and specifically aimed at observing $\nue \rightarrow \nutau$
transitions.

In fact, an emulsion based detector similar to OPERA can perform this
measurement identifying ``wrong-sign'' muons in the magnetic
spectrometer and tracing back the particle up to lead-emulsion or
iron-emulsion bricks to observe the appearance of a decay kink~\cite{silver}.
The pattern of $\nue \rightarrow \nutau$ ($P_{e\tau}$) closely
resembles the one of $\nue \rightarrow \numu$ ($P_{e\mu}$).  If we
expand the $\nu_\mu$ appearance probability to second order in
$\sin 2 \theta_{13}$ and the hierarchy parameter $\alpha \equiv
\sdm/\ldm \simeq
0.03$~\cite{probosc_freund}, we get
\begin{eqnarray}
P_{e\mu} &\simeq& 
\sin^22\theta_{13} \sin^2\theta_{23} 
\frac{\sin^2[(1-\hat{A})\Delta_{31}]}{(1-\hat{A})^2}\nonumber \\
&\pm& \alpha \sin2\theta_{13} \sin\delta \sin2\theta_{12} \sin2\theta_{23}  \sin(\Delta_{31}) \frac{\sin(\hat{A}\Delta_{31})}{\hat{A}}
\frac{\sin[(1-\hat{A})\Delta_{31}]}{(1-\hat{A})} \nonumber \\
&+& \alpha \sin2\theta_{13} \cos\delta \sin2\theta_{12} \sin2\theta_{23}  \cos(\Delta_{31}) \frac{\sin(\hat{A}\Delta_{31})}{\hat{A}}
\frac{\sin[(1-\hat{A})\Delta_{31}]}{(1-\hat{A})} \nonumber \\
&+& \alpha^2 \cos^2\theta_{23} \sin^22\theta_{12} 
\frac{\sin^2(\hat{A}\Delta_{31})}{{\hat{A}}^2} \ ;
\label{equ:papp}
\end{eqnarray}
here, $\Delta_{31} \equiv \ldm L/(4E)$ and $\hat{A} = \pm 2 \,
\sqrt{2} \, E \, G_F \, n_e/\Delta m_{31}^2$, $G_F$ and $n_e$ being
the Fermi constant and the electron density in the earth crust,
respectively. The signs in the second term and $\hat{A}$ are positive
for neutrinos and negative for anti-neutrinos.  On the other hand, in
$P_{e \tau}$ the sign of the second and third terms are flipped and
the replacement $\sin^2 \theta_{23} \leftrightarrow \cos^2
\theta_{23}$ takes place in the first and fourth terms. In particular,
for maximal atmospheric mixing only the signs of the second and third
terms change. As a consequence, the measurement of the ``golden
channel'' $\nue \rightarrow \numu$, of its CP conjugate $\anue
\rightarrow \anumu$ and of the additional transition $\nue \rightarrow
\nutau$ (``silver channel'') can solve the intrinsic degeneracy for
values of \thetaot larger than $1^\circ$ and $\simeq 5$~kton of
detector mass~\cite{silver2}.

More recently, it has been pointed out that a significant improvement
of the Neutrino Factory performance could be achieved relaxing the
energy cut to smaller values and putting up with a worse charge
identification efficiency below
10~GeV~\cite{Cervera:2010rz,Agarwalla:2010hk}. In this region,
however, tau appearance still plays an important
role~\cite{Indumathi:2009hg}. For coarse-grained detectors such as magnetized
iron calorimeters, the golden channel is highly contaminated by
unidentified silver channel events~\cite{Donini:2010xk}, i.e.  $\nue
\rightarrow \nutau$ transitions where the tau decays in $\mu \anumu
\nue$. The detector reconstructs the events as standard ``wrong sign''
muons although the actual neutrino energy shows poor correlation with
the measured muon momentum, due to the large missing energy in $\tau
\rightarrow \mu \anumu \nue$. Hence, the silver channel populates the
golden signal region as a broad-band background and introduces a large
systematic error in the extraction of \thetaot and the CP violating
phase~\cite{Donini:2010xk}. Still, once the silver channel
``background'' is accounted for in the fits of the golden event
spectra, the correct values of the parameters can be
recovered~\cite{Agarwalla:2010hk} and, in addition, the problem of the
degeneracies is further relieved due to the lowering of the muon
energy cut. Clearly, consistency among these spectra, which entangles
subdominant $\nue \rightarrow \nutau$ and
$\nue \rightarrow \numu$ oscillations would be an impressive test for
the standard three-family interpretation of leptonic mixing, even
without an explicit observation of a $\nue \rightarrow \nutau$ sample.

\subsection{$\tau$ appearance and Non Standard Interactions}
\label{sec:future_nsi}

Mass-generation mechanisms for neutrinos naturally produce
perturbations in the Standard Model couplings of these particles, referred to
as non-standard interactions (NSI).
Broadly speaking, NSI can manifest themselves in two different ways.
If they perturb charged-currents they can affect neutrino production and
detection, conversely if they modify neutral-currents they affect
neutrino propagation.

If we consider NSI in the production and detection mechanisms, then the NSI themselves
can be parametrized as a small mixture of the wrong flavor $\nu_\beta$
to a neutrino produced or detected in association with a charged
lepton $\alpha$, i.e.
\begin{equation}
\nu_\alpha^\prime = \nu_\alpha +
\sum_{\beta=e,\mu,\tau}\varepsilon_{\alpha\beta}\nu_\beta\,,
\end{equation}
where the parameters $\varepsilon$ give the strength of the NSI
relative to the standard weak interactions.
Here, a short distance between the
source and the detector is mandatory in order to minimize the
oscillation probability and, therefore, enhance the NSI contribution.
On the other hand, long baseline neutrino experiments mainly constrain
NSI affecting NC since these NSI perturb neutrino oscillation in matter
at macroscopic distances. In this case, NSI appear in the Hamiltonian
describing neutrino propagation in matter:
\begin{equation}
\label{eq:hamiltonian}
\mathcal{H}=\frac{1}{2E}U\left(\begin{array}{ccc}
0&0&0\\
0&\Delta m^2_{21}&0\\
0&0&\Delta m^2_{31}
\end{array}\right)U^\dagger +
\frac{1}{2E}\left(\begin{array}{ccc}
V&0&0\\
0&0&0\\
0&0&0
\end{array}\right)+ \mathcal{H}_{NSI} \,,
\end{equation}
where $U$ is the leptonic mixing matrix and $V=\sqrt{2} G_F n_e$ is
the contribution arising from ordinary matter effects
(Mikheyev-Smirnov-Wolfenstein effect~\cite{MSW}), $G_F$
being the Fermi constant and $n_e$ the density of electrons in the
medium.

 NSI can be simply induced by
non-unitarity of the leptonic mixing matrix - e.g., in see-saw models -
or caused by one-loop exchange of new particles - e.g., by spinless
bosons in supersymmetric models~\cite{Valle:2003uv}.  If the scale of
new physics is high, these effects are small but models where NSI are
significantly enhanced are possible and have been investigated in the
literature. In the 90's the study of these models was boosted by a
possible non-oscillation explanation of the solar neutrino
puzzle~\cite{Guzzo:2001mi}. In recent years, both the hint for a
non-zero value of \thetaot from solar data and the Miniboone
anomalies~\cite{AguilarArevalo:2008rc,AguilarArevalo:2010wv} have
revived the interest on NSI, which can now act as a sub-dominant
contribution to neutrino
oscillations~\cite{GonzalezGarcia:2011my,Palazzo:2011vg,Akhmedov:2010vy};
as a consequence, a precision measurement campaign is considered
mandatory.  Since precision oscillation physics is a young field of
research, limits on NSI affecting the propagation of neutrinos are
rather loose. On the other hand, more stringent bounds can be drawn
from processes that involve the production and detection of neutrinos
in short baseline experiments or radiative contributions to rare
decays~\cite{Davidson:2003ha}.

Even accounting for short baseline experiments, present model
independent bounds on $\varepsilon$ are quite loose -- 
$\mathcal{O}(0.1-1)$. Here, a new generation of high precision experiments at short baselines
would represent the ideal tool to probe much smaller  values of  $\varepsilon$.

A clear evidence for NSI working at production and detection would be
the measurement of a non-unitarity~\cite{Antusch:2006vwa} of the leptonic mixing matrix. A possible
approach to investigate the non-unitarity is to measure all
oscillation probabilities $P(\nu_\mu\rightarrow\nu_\mu),\,
P(\nu_\mu\rightarrow\nu_e),\, P(\nu_\mu\rightarrow\nu_\tau)$ (or
$P(\nu_e\rightarrow\nu_\mu),\, P(\nu_e\rightarrow\nu_e),\,
P(\nu_e\rightarrow\nu_\tau)$) and check that they sum up to one. In this
case the best approach would be to build a dedicated near detector to
measure all oscillation channels.

Indeed, neutrino oscillations constitute an irreducible background for
such possible rare processes. Nevertheless, the short-baseline choice
has the drawback that the NSI signal is given by the product of the
production and detection mechanisms. In order to disentangle the two
mechanisms it is important to exploit different neutrino sources and
as many final states as possible. In this respect, it is
useful to exploit high-energy neutrino beams where the $\tau$
production yield increases with the neutrino energy and, therefore,
final states with $\tau$ leptons can be studied. This consideration
motivated proposals~\cite{Alonso:2010wu} for a new generation of
short-baseline experiments in tau appearance mode. They are aimed at
precisions better than $10^{-6}$ to investigate theory-driven
enhancements of NSI: at these baselines, the dominant contributions come from
the production/detection of neutrinos while propagation effects are
negligible.

If the propagation mechanism is affected by NSI, then long baseline
experiments are necessary. In general, these experiments compete with
the information that can be drawn from current atmospheric
data~\cite{GonzalezGarcia:2011my,Maltoni:2008mu}, but the purity of
the source can be fruitfully employed to tighten some bounds. In
particular, tau appearance in OPERA can be used to improve our
knowledge of $\epsilon_{\mu \tau}$~\cite{Blennow:2008ym}~\footnote{For
a definition of the $\epsilon$ couplings relevant for neutrino
propagation, see e.g.~\cite{GonzalezGarcia:2011my}.} while the CNGS
statistics is too poor to impact bounds on $\epsilon_{\tau \tau}$ or
$\epsilon_{e \tau}$~\cite{EstebanPretel:2008qi}. Large statistics tau
appearance experiments where matter effects are dominant and,
therefore, NSI due to propagation are sizable, are possible only in
the framework of the Neutrino Factories (see
Sec.~\ref{sec:future_standard}).  A systematic assessment of
performance of these facilities with respect to NSI has been carried
out in~\cite{Kopp:2008ds}. Here, the use of the tau appearance
channel, the ``silver channel'' of Sec.~\ref{sec:future_standard}, is
relevant only for high energy neutrino factories with muon energies
larger than $\sim$25~GeV and its sensitivity is mainly limited to
$\epsilon_{e\tau}$.

\subsection{$\tau$ appearance and sterile neutrinos}

The possible confirmation of the LSND anomaly~\cite{Aguilar:2001ty} by
the MiniBoone antineutrino data~\cite{AguilarArevalo:2010wv} makes the
study of hypothetical neutrinos that are singlet under the electroweak
gauge groups, sterile neutrinos, a very lively field of research.  It
is, therefore, interesting to assess the contribution of the tau
appearance channel to the clarification of this issue, at least in the
simplest scenarios where just one sterile neutrino is added to the
three active families, the 3+1 scheme. Unfortunately, this
oversimplified scheme is not able to account for all experimental data
and the current global fit is very poor. More sophisticated models
(see e.g.,\cite{Kopp:2011qd}) show better agreement with data but the
3+1 scheme illustrates very well the experimental
challenges. Moreover, many of the considerations made in
Sec.~\ref{sec:future_nsi} hold for sterile neutrino searches, too.
The impact of $\nu_\mu\rightarrow\nu_\tau$ appearance searches in the
framework of the 3+1 scheme has been studied for both
conventional~\cite{Donini:2007yf} and Neutrino Factory
beams~\cite{Donini:2008wz,Meloni:2010zr}. In the first case, if the
OPERA detector is exposed to the nominal CNGS beam intensity, a null
result can slightly improve the present bound on
$\theta_{13}$~\footnote{This angle is different from the \thetaot
angle defined in Sec.\ref{sec:intro}. It represents the mixing angle
between the first and third family in a 3 active + 1 sterile neutrino
mixing scheme. }  but not those on the active-sterile mixing angles,
$\theta_{14}$, $\theta_{24}$ and $\theta_{34}$. If the beam intensity
is increased by a factor 2 or beyond, not only the sensitivity to
$\theta_{13}$ increases accordingly, but a significant sensitivity to
$\theta_{24}$ and $\theta_{34}$ can be achieved. The $\theta_{24}$ and
$\theta_{34}$ sensitivities strongly depend on the value of the
CP-violating phase $\delta_3$, highest sensitivities being available
for $\delta_3 \simeq \pi/2$. In order to reach significant
improvements on $\theta_{13}$, the angle should better be constrained
by high-intensity $\nu_e$ disappearance experiments. Once more (see
Sec.~\ref{sec:future_nsi}), OPERA is limited by the small detector
mass.  It is, however, very interesting to note that the sensitivity
of OPERA to \thetaot and to the other angles of the 3+1 model mainly
comes from the study of \numunutau\ transitions, while the
corresponding sensitivity due to \numunue , the CP conjugate of the
LSND measurement, is marginal. This is due to the rather large
baseline compared with LSND/Miniboone and to the additional
constraints coming from the SuperKamiokande atmospheric data.

Clearly, the results that can be obtained at a Neutrino Factory are
much better than those obtained by exploiting the CNGS beam,
even assuming a major upgrade of the
facility~\cite{Donini:2007yf}. As for the case of the silver channel
(Sec.~\ref{sec:future_standard}), the setup must be equipped with a
massive OPERA-like detector; the ideal baseline is, nevertheless,
$\sim3000$~km, i.e. the detector can be positioned in the same
underground site as for the magnetized calorimeter.  In this case, the
detector seeks for ``right-sign'' muons in coincidence with a decay
kink, i.e. it measures the leading \numunutau\ transition, which is sometimes
called the ``discovery channel'', in contrast with the \nuenutau\ silver
channel of Sec.~\ref{sec:future_standard}.  Further improvements
can be achieved exploiting Magnetized Emulsion Cloud
Chambers~\cite{Abe:2007bi,Fukushima:2008zzb} made of iron bricks and
photographic films. The magnetization of the iron allows for
sign measurement of the final state particles even in the
occurrence of muon-less tau decays. A viable alternative to the study
of tau appearance seems, however, the exploitation of near detectors
located close to the muon decay ring~\cite{Meloni:2010zr}, especially
if $\Delta m_{41} \gg \Delta m_{31}$.

\section{Conclusions}
The search for an explicit observation of flavor changing neutrino
oscillations by identifying a different lepton than the one of the
initial flavor has a decades long history. In 1998 it became clear
that the bulk of oscillations at the atmospheric scale is likely
constituted by \numunutau\ transitions: this consideration has boosted
enormously the search for tau appearance in long-baseline experiments
both with natural and artificial sources. Between 2006 and 2010,
SuperKamiokande and OPERA have gained evidence of such transition using
quite different techniques and significant improvements are expected
in the next few years. Although no dedicated facilities for precision
measurements of tau appearance have been designed so far, it is clear
that the study of \numunutau\ will play a relevant role in any
experiment operating beyond the kinematical threshold for tau
production and especially in the far detectors of the Neutrino
Factories. Novel short-baseline experiments along the line of CHORUS
and NOMAD can be of interest beyond the standard three-family
oscillation scenario, mainly for precision searches of non-standard
interactions.

\section*{Acknowledgments}

We wish to express our gratitude to Andrea Donini, Antonio Ereditato
and Chris Walter for many useful discussions and careful reading
of the manuscript.


\section*{References}
\begin{thebibliography}{10}
\bibitem{giunti_book}
  C.~Giunti, C.~W.~Kim,
  ``Fundamentals of Neutrino Physics and Astrophysics,'', Oxford University Press,
  Oxford, UK, 2007.
\bibitem{pontecorvo} B.~Pontecorvo, Sov.\ Phys.\ JETP {\bf 6} (1957)
429 [Zh.\ Eksp.\ Teor.\ Fiz.\ {\bf 33} (1957) 549]; B.~Pontecorvo,
Sov.\ Phys.\ JETP {\bf 6} (1968) 984 [Zh.\ Eksp.\ Teor.\ Fiz.\ {\bf
53} (1967) 1717].
\bibitem{neutrino_osc}
Y. Katayama, K. Matunoto, S. Tanaka and E. Yamada, Progr. Theor. Phys. {\bf 28} (1962) 675;
Z.~Maki, M.~Nakagawa and S.~Sakata,
Prog.\ Theor.\ Phys.\  {\bf 28} (1962) 870;
B.~Pontecorvo,
Sov.\ Phys.\ JETP {\bf 26} (1968) 984 [Zh.\ Eksp.\ Teor.\ Fiz.\  {\bf 53} (1967) 1717];
V.~N.~Gribov and B.~Pontecorvo,
Phys.\ Lett.\ B {\bf 28} (1969) 493.

\bibitem{SchwetzTortolaValle}
 T.~Schwetz, M.~A.~Tortola and J.~W.~F.~Valle,
  New J.\ Phys.\  {\bf 10} (2008) 113011.

\bibitem{minos_adamson} P.~Adamson {\it et al.} [ The MINOS Collaboration ],
  ``Measurement of the neutrino mass splitting and flavor mixing by MINOS,'' arXiv:1103.0340 [hep-ex].

\bibitem{kamland}
S.~Abe {\it et al.}  [KamLAND Collaboration],
  Phys.\ Rev.\ Lett.\  {\bf 100} (2008) 221803.

\bibitem{K2K}
M.H.~Ahn {\it et al.}  [K2K Coll.], Phys.\ Rev.\ Lett.\
{\bf 90} (2003) 041801;
M.~H.~Ahn {\it et al.}  [K2K Collaboration],
  Phys.\ Rev.\  D {\bf 74} (2006) 072003.

\bibitem{minos}
D.~G.~Michael {\it et al.} [ MINOS Collaboration ],
Phys.\ Rev.\ Lett.\  {\bf 97 } (2006)  191801.
P.~Adamson {\it et al.} [ MINOS Collaboration ],
Phys.\ Rev.\  {\bf D77 } (2008)  072002;
P.~Adamson {\it et al.} [ MINOS Collaboration ],
Phys.\ Rev.\ Lett.\  {\bf 101 } (2008)  131802.


\bibitem{chooz}
M.~Apollonio {\it et al.}  [ CHOOZ Collaboration ],
Eur.\ Phys.\ J.\ C {\bf 27} (2003) 331.
\bibitem{palo_verde}
F.~Boehm {\it et al.} [ Palo Verde Collaboration ],
Phys.\ Rev.\  D {\bf 64} (2001) 112001.
\bibitem{Mezzetto:2010zi}
M.~Mezzetto, T.~Schwetz,
J.\ Phys.\ {\bf G37 } (2010)  103001.
\bibitem{PDG} K. Nakamura, S. T. Petcov ``Neutrino mass, mixing, and
oscillations'' in K. Nakamura et al. (Particle Data Group), J. Phys. G
37, 075021 (2010).
\bibitem{tribimaximal} P.F. Harrison, D.H. Perkins and W.G. Scott, Phys. Lett. B {\bf 530} (2002) 167.


\bibitem{Harari:1988ew}
 H.~Harari,
 Phys.\ Lett.\  {\bf B216 } (1989)  413.

\bibitem{Ellis:1992zr}
 J.~R.~Ellis, J.~L.~Lopez, D.~V.~Nanopoulos,
 Phys.\ Lett.\  {\bf B292 } (1992)  189.

\bibitem{seesaw}
P. Minkowski, Phys. Lett. B {\bf 67}  (1977) 421; T. Yanagida, in Proceedings of the Workshop
on the Unified Theory and the Baryon Number in the Universe (O. Sawada
and A. Sugamoto, eds.), KEK, Tsukuba, Japan, 1979, p. 95; M. Gell-Mann, P. Ramond,
and R. Slansky,  in Supergravity (P. van
Nieuwenhuizen and D. Z. Freedman, eds.), North Holland, Amsterdam, 1979, p. 315;
S. L. Glashow, in Proceedings of the
1979 Cargese Summer Institute on Quarks and Leptons, Plenum Press, New York,
1980, p. 687; R. N. Mohapatra and G. Senjanovic, Phys. Rev. Lett. {\bf 44} (1980) 912.

\bibitem{Altegoer:1997gv}
 J.~Altegoer {\it et al.} [ NOMAD Collaboration ],
 Nucl.\ Instrum.\ Meth.\  {\bf A404 } (1998)  96.

\bibitem{Eskut:1997ar}
 E.~Eskut {\it et al.} [ CHORUS Collaboration ],
 Nucl.\ Instrum.\ Meth.\  {\bf A401 } (1997)  7.

\bibitem{wanf}
E. Heijne. CERN Yellow Report 83-06  (1983); G. Acquistapace {\it et al.} CERN Preprint CERN-ECP/95-14 (1995);
L. Casagrande {\it et al.} CERN Yellow Report 96-06 (1996).

\bibitem{collaboration:2010hi}
 N.~Abgrall {\it et al.},
``Time Projection Chambers for the T2K Near Detectors,''
arXiv:1012.0865 [physics.ins-det].

\bibitem{Ushida:1986zn}
 N.~Ushida {\it et al.} [ FERMILAB E531 Collaboration ],
 Phys.\ Rev.\ Lett.\  {\bf 57 } (1986)  2897.

\bibitem{McFarland:1995sr}
 K.~S.~McFarland, D.~Naples, C.~G.~Arroyo, P.~S.~Auchincloss, P.~de Barbaro, A.~O.~Bazarko, R.~H.~Bernstein, A.~Bodek {\it et al.},
 Phys.\ Rev.\ Lett.\  {\bf 75 } (1995)  3993.


\bibitem{emul}
 S.~Aoki, E.~Barbuto, C.~Bozza, J.~P.~Fabre, W.~Flegel, G.~Grella, M.~Guler, T.~Hara {\it et al.},
 Nucl.\ Instrum.\ Meth.\  {\bf A447 } (2000)  361.

\bibitem{fiber}
P.~Annis {\it et al.}  [ CHORUS Collaboration],
Nucl.\ Instrum.\ Meth.\ A {\bf 412} (1998) 19.

\bibitem{hex}
F.~Bergsma {\it et al.},
Nucl.\ Instrum.\ Meth.\ A {\bf 357} (1995) 243.

\bibitem{hcc}
J.~W.~E.~Uiterwijk {\it et al.},
Nucl.\ Instrum.\ Meth.\ A {\bf 409} (1998) 682.

\bibitem{calo}
E.~Di Capua {\it et al.},
Nucl.\ Instrum.\ Meth.\ A {\bf 378} (1996) 221.

\bibitem{Eskut:2007rn}
 E.~Eskut {\it et al.} [ CHORUS Collaboration ],
 Nucl.\ Phys.\  {\bf B793 } (2008)  326.

\bibitem{Astier:2001yj}
 P.~Astier {\it et al.} [ NOMAD Collaboration ],
 Nucl.\ Phys.\  {\bf B611 } (2001)  3.



\bibitem{sk_1998} Y.~Fukuda {\it et al.}  [Super-Kamiokande
Coll.], Phys.\ Rev.\ Lett.\ {\bf 81} (1998) 1562.
\bibitem{atm}
Y.~Ashie {\it et al.}  [Super-Kamiokande Coll.],
Phys. Rev. Lett. {\bf 93} (2004) 101801;
M.~Ambrosio {\it et al.}  [MACRO Coll.], Phys.\ Lett.\ B {\bf 517}
(2001) 59; 
M.~Sanchez {\it et al.}  [Soudan 2 Coll.],
Phys.\ Rev.\ D {\bf 68} (2003) 113004.


\bibitem{sterile} S.~Fukuda {\it et al.} [ Super-Kamiokande Collaboration ], 
Phys.\ Rev.\ Lett.\ {\bf 85 } (2000) 3999; 
P.~Adamson {\it et al.} [ MINOS Collaboration ],
Phys.\ Rev.\  Lett.\ {\bf 101 } (2008) 221804.
\bibitem{sk_pattern} 
Y.~Ashie {\it et al.} [ Super-Kamiokande Collaboration ],
Phys.\ Rev.\ Lett.\  {\bf 93 } (2004)  101801.
\bibitem{pasquali}  L.~Pasquali, M.~H.~Reno,
Phys.\ Rev.\  {\bf D59 } (1999)  093003.
\bibitem{sk_app}
K.~Abe {\it et al.} [ Super-Kamiokande Collaboration ],
Phys.\ Rev.\ Lett.\  {\bf 97 } (2006)  171801.
\bibitem{lar} C.~Rubbia, ``The liquid Argon Time Projection
Chamber: a new concept for Neutrino Detector, CERN-EP/77-08.
\bibitem{600t}
S.~Amerio {\it et al.}  [ ICARUS Collaboration ],
Nucl.\ Instrum.\ Meth.\  A {\bf 527} (2004) 329;
A.~Menegolli [ ICARUS Collaboration ],
J.\ Phys.\ Conf.\ Ser.\  {\bf 203 } (2010)  012107.
\bibitem{Conrad:2010mh}
J.~Conrad, A.~de Gouvea, S.~Shalgar {\it et al.},
Phys.\ Rev.\  {\bf D82 } (2010)  093012.
\bibitem{Schwetz:2008er}
T.~Schwetz, M.~A.~Tortola, J.~W.~F.~Valle,
New J.\ Phys.\  {\bf 10 } (2008)  113011.
\bibitem{doublechooz}
F.~Ardellier {\it et al.}  [ Double Chooz Collaboration ],
arXiv:hep-ex/0606025.
\bibitem{reno}
S.~B.~Kim  [ RENO Collaboration ],
AIP Conf.\ Proc.\  {\bf 981} (2008) 205
[J.\ Phys.\ Conf.\ Ser.\  {\bf 120} (2008) 052025].


\bibitem{t2k}
Y.~Itow {\it et al.}  [ T2K Collaboration ],
arXiv:hep-ex/0106019.
\bibitem{dayabay}
X.~Guo {\it et al.}  [ Daya-Bay Collaboration ],
arXiv:hep-ex/0701029.
\bibitem{nova}
D.~S.~Ayres {\it et al.}  [ NOvA Collaboration ],
arXiv:hep-ex/0503053. \\
See also \texttt{http://www-nova.fnal.gov}.

\bibitem{xsect} Y.~Hayato, Nucl. Phys. B (Proc. Suppl.) {\bf 112}
(2002) 171; K. Hagiwara et al., Nucl. Phys. B {\bf 668} (2003) 364.
\bibitem{Itow:2008zza}
Y.~Itow [ Super-Kamiokande Collaboration ],
J.\ Phys.\ Conf.\ Ser.\  {\bf 120 } (2008)  052037.
\bibitem{kato_thesis} T.~Kato, Ph.D. Thesis, Stony Brook University, May 2007 (available
at \texttt{http://www-sk.icrr.u-tokyo.ac.jp/sk/pub/index.html}).

\bibitem{nnn2010} R.~Wendell, Talk at 11th International Workshop on
Next generation Nucleon Decay and Neutrino Detectors, December 13-16, 2010, Toyama, Japan.



\bibitem{CNGS} Ed. K. Elsener, ``The CERN Neutrino beam to Gran Sasso
(Conceptual Technical Design)'', CERN 98-02, INFN/AE-98/05; R. Bailey
et al., ``The CERN Neutrino beam to Gran Sasso (NGS)'' (Addendum to
report CERN 98-02, INFN/AE-98/05), CERN-SL/99-034(DI), INFN/AE-99/05.

\bibitem{operaold}
A. Ereditato, K. Niwa and P. Strolin, The emulsion technique for short, medium and long baseline $\nu_\mu\rightarrow\nu_\tau$ oscillation experiments, 423, INFN-AE-97-06, DAPNU-97-07;\\
OPERA collaboration, H. Shibuya et al., Letter of intent: the OPERA emulsion detector for a long-baseline neutrino-oscillation experiment, CERN-SPSC-97-24, LNGS-LOI-8-97.

\bibitem{proposal1}
M. Guler {\it et al.} [ OPERA collaboration ], An appearance experiment to search for $\nu_\mu\rightarrow\nu_\tau$ oscillations in the CNGS beam: experimental proposal, CERN-SPSC-2000-028,  LNGS P25/2000

\bibitem{proposal2}
M. Guler et al.,  [ OPERA collaboration ] Status Report on the OPERA experiment, CERN/SPSC 2001-025, LNGS-EXP 30/2001 add. 1/01

\bibitem{operafirst} R. Acquafredda {\it et al.}  [ OPERA
collaboration ], New J. Phys. {\bf 8} (2006) 303; 
N. Agafonova {\it et al.}  [ OPERA collaboration ], JINST {\bf 4} (2009) P06020.

\bibitem{donut}
  K.~Kodama {\it et al.} [ DONUT Collaboration ],
  Phys.\ Lett.\  {\bf B504 } (2001)  218;
K.~Kodama {\it et al.} [ DONuT Collaboration ],
  Phys.\ Rev.\  {\bf D78 } (2008)  052002.


\bibitem{operadetector}
R. Acquafredda {\it et al.}  [ OPERA collaboration ], JINST {\bf 4} (2009) P04018.

\bibitem{ESS}
N. Armenise {\it et al.}, Nucl. Instrum. Meth. {\bf A551} (2005) 261; M. De Serio {\it et al.}, Nucl. Instrum. Meth. {\bf A554} (2005) 247;
L. Arrabito {\it et al.},  Nucl. Instrum. Meth. {\bf A568} (2006) 578.

\bibitem{SUTS}
K. Morishima and T. Nakano, JINST (2010) 5 P04011.

\bibitem{facilities} A. Anokhina {\it et al.}  [ OPERA collaboration
], JINST {\bf 3} (2008) P07002; A. Anokhina {\it et al.}  [ OPERA
collaboration ], JINST {\bf 3} (2008) P07005; T. Adam {\it et al.} [ OPERA
collaboration ], Nucl. Instrum. Meth. {\bf A577} (2007) 523; T. Nakamura
{\it et al.}  [ OPERA collaboration ], Nucl. Instrum. Meth. {\bf A556}
(2006) 80.




\bibitem{Agafonova:2010dc}
 N.~Agafonova {\it et al.} [ OPERA Collaboration ],
 Phys.\ Lett.\  {\bf B691 } (2010)  138.



\bibitem{nufact}
S.~Geer, Phys.\ Rev.\ D {\bf 57} (1998) 6989
[Erratum-ibid.\ D {\bf 59} (1999) 039903];
A.~De Rujula, M.~B.~Gavela and P.~Hernandez,
Nucl.\ Phys.\  B {\bf 547} (1999) 21.
\bibitem{veryhigheBB}
F.~Terranova,
A.~Marotta, P.~Migliozzi and M.~Spinetti, Eur.\ Phys.\ J.\ C {\bf 38}
(2004) 69; S. K. Agarwalla, A. Raychaudhuri and A. Samanta,
Phys. Lett. B {\bf 629} (2005) 33;  
S.~K.~Agarwalla, S.~Choubey, A.~Raychaudhuri {\it et al.},
JHEP {\bf 0806 } (2008)  090;
S.~K.~Agarwalla, S.~Choubey, A.~Raychaudhuri,
Nucl.\ Phys.\  {\bf B805 } (2008)  305.

\bibitem{MIND}
A.~Cervera-Villanueva,
``ISS/IDS detector study,''
AIP Conf.\ Proc.\  {\bf 981} (2008) 51.
\bibitem{ISS_phys}
A.~Bandyopadhyay {\it et al.}  [ISS Physics Working Group],
arXiv:0710.4947 [hep-ph].
\bibitem{ISS_acc}
J.~S.~Berg {\it et al.}  [ISS Accelerator Working Group],
JINST {\bf 4} (2009) P07001.


\bibitem{intrinsic_deg}
J.~Burguet-Castell, M.~B.~Gavela, J.~J.~Gomez-Cadenas, P.~Hernandez and O.~Mena,
Nucl.\ Phys.\ B {\bf 608} (2001) 301; 
H.~Minakata and H.~Nunokawa,
JHEP {\bf 0110}  (2001) 001;
V.~Barger, D.~Marfatia and K.~Whisnant,
Phys.\ Rev.\ D {\bf 65}  (2002) 073023.
\bibitem{betabeam}
P.~Zucchelli,
Phys.\ Lett.\ B {\bf 532}  (2002) 166;
M.~Lindroos and M.~Mezzetto, ``Artificial Neutrino Beams: Beta Beams'',
Imperial College Press, Aug. 2009.


\bibitem{silver}
A.~Donini, D.~Meloni and P.~Migliozzi, Nucl.\ Phys.\ B {\bf
646} (2002) 321. 
\bibitem{probosc_freund}
A.~Cervera {\it et al.}, Nucl. Phys. {\bf B579}~(2000)~17, erratum ibid.
Nucl. Phys. {\bf B593}~(2001)~731; 
M.~Freund, Phys. Rev.~{\bf D64}~(2001)~053003;
E.~K.~Akhmedov, R.~Johansson, M.~Lindner {\it et al.},
JHEP {\bf 0404 } (2004)  078.
\bibitem{silver2}
D.~Autiero {\it et al.}, Eur.\ Phys.\ J.\ {\bf C33} (2004) 243.
\bibitem{Cervera:2010rz}
A.~Cervera, A.~Laing, J.~Martin-Albo {\it et al.},
Nucl.\ Instrum.\ Meth.\  {\bf A624 } (2010)  601.
\bibitem{Agarwalla:2010hk}
S.~K.~Agarwalla, P.~Huber, J.~Tang {\it et al.},
JHEP {\bf 1101 } (2011)  120.
\bibitem{Indumathi:2009hg}
D.~Indumathi, N.~Sinha,
Phys.\ Rev.\  {\bf D80 } (2009)  113012.
\bibitem{Donini:2010xk}
A.~Donini, J.~J.~Gomez Cadenas, D.~Meloni,  arXiv:1005.2275 [hep-ph].
\bibitem{MSW}
L. Wolfenstein, Phys. Rev. {\bf D17} (1978) 2369;
S. P. Mikheyev and A. Y. Smirnov, Sov. J. Nucl. Phys. {\bf 42}  (1985) 913;
S. P. Mikheyev and A. Y. Smirnov, Nuovo Cim. {\bf C9}  (1986) 17.
\bibitem{Valle:2003uv}
 J.~W.~F.~Valle,
 J.\ Phys. {\bf G29 } (2003)  1819.
\bibitem{Guzzo:2001mi}
 M.~Guzzo, P.~C.~de Holanda, M.~Maltoni, H.~Nunokawa, M.~A.~Tortola, J.~W.~F.~Valle,
 Nucl.\ Phys.\  {\bf B629 } (2002)  479.
\bibitem{AguilarArevalo:2008rc}
 A.~A.~Aguilar-Arevalo {\it et al.} [ MiniBooNE Collaboration ],
 Phys.\ Rev.\ Lett.\  {\bf 102 } (2009)  101802.
\bibitem{AguilarArevalo:2010wv}
 A.~A.~Aguilar-Arevalo {\it et al.} [ MiniBooNE Collaboration ],
 Phys.\ Rev.\ Lett.\  {\bf 105 } (2010)  181801.
\bibitem{GonzalezGarcia:2011my}
  M.~C.~Gonzalez-Garcia, M.~Maltoni, J.~Salvado,
  JHEP {\bf 1105 } (2011)  075.
%
\bibitem{Palazzo:2011vg}
 A.~Palazzo,
  Phys.\ Rev.\  {\bf D83 } (2011)  101701.
\bibitem{Akhmedov:2010vy}
 E.~Akhmedov, T.~Schwetz,
 JHEP {\bf 1010 } (2010)  115.
\bibitem{Davidson:2003ha}
 S.~Davidson, C.~Pena-Garay, N.~Rius, A.~Santamaria,
 JHEP {\bf 0303 } (2003)  011.
\bibitem{Antusch:2006vwa}
  S.~Antusch, C.~Biggio, E.~Fernandez-Martinez, M.~B.~Gavela, J.~Lopez-Pavon,
  JHEP {\bf 0610 } (2006)  084.
\bibitem{Alonso:2010wu} R.~Alonso, S.~Antusch, M.~Blennow, P.~Coloma,
 A.~de Gouvea, E.~Fernandez-Martinez, B.~Gavela, C.~Gonzalez-Garcia
 {\it et al.}, ``Summary report of MINSIS workshop in Madrid,'',
 arXiv:1009.0476 [hep-ph].
\bibitem{Maltoni:2008mu}
 M.~Maltoni,
 J.\ Phys.\ Conf.\ Ser.\  {\bf 136 } (2008)  022024.
\bibitem{Blennow:2008ym}
 M.~Blennow, D.~Meloni, T.~Ohlsson, F.~Terranova, M.~Westerberg,
 Eur.\ Phys.\ J.\  {\bf C56 } (2008)  529.
\bibitem{EstebanPretel:2008qi}
 A.~Esteban-Pretel, J.~W.~F.~Valle, P.~Huber,
 Phys.\ Lett.\  {\bf B668 } (2008)  197.
\bibitem{Kopp:2008ds}
 J.~Kopp, T.~Ota, W.~Winter,
 Phys.\ Rev.\  {\bf D78 } (2008)  053007.
\bibitem{Aguilar:2001ty}
 A.~Aguilar {\it et al.} [ LSND Collaboration ],
 Phys.\ Rev.\  {\bf D64 } (2001)  112007.
\bibitem{Kopp:2011qd}
 J.~Kopp, M.~Maltoni, T.~Schwetz,
 ``Are there sterile neutrinos at the eV scale?,''  
 arXiv:1103.4570 [hep-ph].
\bibitem{Donini:2007yf}
 A.~Donini, M.~Maltoni, D.~Meloni, P.~Migliozzi, F.~Terranova,
 JHEP {\bf 0712 } (2007)  013.
\bibitem{Donini:2008wz}
 A.~Donini, K.~Fuki, J.~Lopez-Pavon, D.~Meloni, O.~Yasuda,
 JHEP {\bf 0908 } (2009)  041.
\bibitem{Meloni:2010zr}
D.~Meloni, J.~Tang, W.~Winter,
 Phys.\ Rev.\  {\bf D82 } (2010)  093008.
\bibitem{Abe:2007bi}
 T.~Abe {\it et al.} [ ISS Detector Working Group Collaboration ],
 JINST {\bf 4 } (2009)  T05001.
\bibitem{Fukushima:2008zzb}
 C.~Fukushima, M.~Kimura, S.~Ogawa, H.~Shibuya, G.~Takahashi, K.~Kodama, T.~Hara, S.~Mikado,
 Nucl.\ Instrum.\ Meth.\  {\bf A592 } (2008)  56.




\end{thebibliography}
\end{document}